\documentclass[12pt,preprint]{aastex}	
\newcommand {\apl}	{{\it Ap.\ Lett.}}
\newcommand {\ban}	{{\it Bull.\ Astr.\ Inst.\ Netherlands}}
\newcommand {\jkas}	{{\it J.\ Korean Ast.\ Soc.}}

\newcommand {\rvmx}	{{\it Rev.\ Mex.\ Astron.\ Astrofis.}}

\newcommand {\kms}	{{\mbox{$\rm \,km\,s^{-1}$}}}
\newcommand {\kkms}	{{\mbox{$\rm \,K\,km\,s^{-1}$}}}
\newcommand {\kdeg}	{{\mbox{$\rm \,K\,deg$}}}
\newcommand {\kddeg}	{{\mbox{$\rm \,K\,deg^2$}}}
\newcommand {\ddeg}	{{\mbox{$\rm \,deg^2$}}}
\newcommand {\msun}	{{\mbox{$\rm \,M_\odot$}}}
\newcommand {\hii}	{{\ion{H}{2}}}
\newcommand {\hi}	{{\ion{H}{1}}}
\newcommand {\hh}	{{\mbox{H$_2$}}}

\newcommand {\twco}	{{\mbox{$^{12}$CO}}}

\newcommand {\perarea}	{{\mbox{$\rm \,cm^{-2}$}}}

\newcommand {\lbvunit}	{{\mbox{$\rm \,deg^2\,km\,s^{-1}$}}}
\newcommand {\absintunit}{{\mbox{$\rm \,K\,deg^2\,km\,s^{-1}$}}}
\newcommand {\scinot}[2] {#1\times10^{#2}}
\newcommand {\subhisa}	{{_{_{\!H\!I\!S\!A}}}}

\newcommand {\subon}	{{_{_{\!O\!N}}}}
\newcommand {\suboff}	{{_{_{\!O\!F\!F}}}}
\newcommand {\subunabs}	{{_{_{\!U}}}}
\newcommand {\subcont}	{{_{_{\!C}}}}
\newcommand {\subspin}	{{_{_{\!S}}}}

\newcommand {\subrms}	{{_{\!rms}}}

\newcommand {\sublsr}	{{_{_{\!L\!S\!R}}}}

\newcommand {\glon}	{{\ell}}
\newcommand {\glat}	{{b}}
\newcommand {\vlsr}	{{v\sublsr}}
\newcommand {\vel}	{{v}}

\newcommand {\dec}	{{\delta}}

\newcommand {\ton}	{{T\subon}}
\newcommand {\toff}	{{T\suboff}}
\newcommand {\tu}	{{T\subunabs}}
\newcommand {\tc}	{{T\subcont}}
\newcommand {\ts}	{{T\subspin}}

\newcommand {\trms}	{{T\subrms}}

\newcommand {\dt}	{{\Delta T}}
\newcommand {\dv}	{{\Delta v}}
\newcommand {\da}	{{\Delta \theta}}

\newcommand {\bfrac}	{{p}}

\newcommand {\coldens}	{{N\subhisa}}

\newcommand {\nvox}	{{N_{voxels}}}

\newcommand {\wmos}	{{w_m}}

\newcommand {\fid}	{{f_{det}}}
\newcommand {\fot}	{{f_{true}}}
\newcommand {\fotfracvol}{{\eta_{V}}}
\newcommand {\fotfracint}{{\eta_{\dt}}}
\newcommand {\tufrac}	{{\phi_{U}}}

\newcommand {\tuavg}	{{\langle \tu \rangle}}
\newcommand {\dtavg}	{{\langle \dt \rangle}}
\newcommand {\dti}	{{\dt_{in}}}
\newcommand {\tui}	{{\tu_{in}}}
\newcommand {\dai}	{{\da_{in}}}
\newcommand {\dvi}	{{\dv_{in}}}

\newcommand {\tuo}	{{\tu_{out}}}

\newcommand {\ddt}      {{\Delta\dt}}
\newcommand {\dtu}	{{\Delta\tu}}
\newcommand {\dda}	{{\Delta\da}}
\newcommand {\ddv}	{{\Delta\dv}}

\shorttitle{A HISA Census of the CGPS}
\shortauthors{Gibson et al.}

\begin{document}

\title{A Self-Absorption Census of Cold \hi\ Clouds 
	in the Canadian Galactic Plane Survey}

\author{Steven J. Gibson\altaffilmark{1}, 
	A. Russell Taylor\altaffilmark{1},
	Lloyd A. Higgs\altaffilmark{2},
	Christopher M. Brunt\altaffilmark{3},
	and Peter E. Dewdney\altaffilmark{2}}

\altaffiltext{1}{Dept. of Physics \& Astronomy, University of Calgary, 
2500 University Drive N.W., Calgary, Alberta T2N~1N4, Canada;
gibson@ras.ucalgary.ca; russ@ras.ucalgary.ca}

\altaffiltext{2}{Dominion Radio Astrophysical Observatory, 
Box 248, Penticton, British Columbia V2A~6K3, Canada;
Lloyd.Higgs@nrc.ca; Peter.Dewdney@nrc.ca}

\altaffiltext{3}{Department of Astronomy, LGRT 632, University of
Massachusetts, 710 North Pleasant Street, Amherst, MA 01003;
brunt@roobarb.astro.umass.edu
\vskip 0.35in
{\large \bf
\noindent For a version of this paper with full-resolution figures, please see 
\hspace*{0.35in} http://www.ras.ucalgary.ca/$\sim$gibson/hisa/cgps1\_survey/.
}
}

\begin{abstract}

We present a 21cm line \hi\ self-absorption (HISA) survey of cold atomic gas
within Galactic longitudes $\glon = 75\arcdeg$ to $146\arcdeg$ and latitudes
$\glat = -3\arcdeg$ to $+5\arcdeg$.  
We identify HISA as spatially and spectrally confined dark \hi\ features and
extract it from the surrounding \hi\ emission in the arcminute-resolution
Canadian Galactic Plane Survey (CGPS).  We compile a catalog of the most
significant features in our survey and compare our detections against those in
the literature.  Within the parameters of our search, we find nearly all
previously detected features and identify many new ones.  The CGPS shows HISA
in much greater detail than any prior survey and allows both new and
previously-discovered features to be placed into the larger context of Galactic
structure.  In space and radial velocity, faint HISA is detected virtually
everywhere that the \hi\ emission background is sufficiently bright.  This
ambient HISA population may arise from small turbulent fluctuations of
temperature and velocity in the neutral interstellar medium.  By contrast,
stronger HISA is organized into discrete complexes, many of which follow a
longitude-velocity distribution that suggests they have been made visible by
the velocity reversal of the Perseus arm's spiral density wave.  The cold \hi\
revealed in this way may have recently passed through the spiral shock and be
on its way to forming molecules and, eventually, new stars.  This paper is the
second in a series examining HISA at high angular resolution.  A companion
paper (Paper~III) describes our HISA search and extraction algorithms in
detail.

\end{abstract}


\section{Introduction}
\label{Sec:intro}

\subsection{Mapping Cold Galactic \hi}

The cold ($T \la 100$~K) phase of neutral atomic hydrogen gas (\hi) is an
important probe of the mass, structure, and evolutionary state of the Galactic
interstellar medium (ISM).  Although it occupies only a small fraction of the
ISM volume \citep{kh88}, cold \hi\ contains $\sim 30$\% of the total gas mass
near the Sun \citep{w03}.  On sub-parsec scales, cold \hi\ clouds display
significant structure (\citealt{g00}, hereafter Paper~I; \citealt{b05}) that
may be shaped by shocks, turbulence, or magnetic fields.  On kiloparsec scales,
the cold \hi\ distribution may trace spiral structure in the Galactic disk
\citep{g02,g04}.  Cold \hi\ may even offer a window into the process of
molecular condensation.  It is often found with molecular gas in the quiescent
environments needed for star formation \citep{k74}, and it can be used to study
turbulent diffusion \citep*{wla02} and cosmic-ray dissociation \citep{lg03}
inside molecular clouds.  But cold \hi\ without apparent molecular gas is also
found (\citealt{gd89}; Paper~I), sometimes at temperatures too cold for
standard equilibrium ISM models to explain easily \citep{w03}; such gas may be
cooling and forming \hh.

Investigating these issues thoroughly requires mapping cold \hi\ over large
areas, but this is not straightforward.  The principal \hi\ tracer, 21cm line
emission, does not show cold gas conspicuously, since the line brightness is
independent of temperature when optically thin and proportional to temperature
when optically thick \citep{g02}.  In fact, since \hi\ column densities
computed from 21cm line emission presume optically thin conditions, they miss
much of the coldest, densest \hi\ and underestimate the total mass.  However,
the inverse dependence of optical depth on temperature allows cold \hi\ to be
readily examined in {\it absorption}.  \hi\ continuum absorption (HICA) studies
find that typically 30\% of the total \hi\ mass is missed by emission columns
in the Galactic plane \citep{d03,st04}, with almost 50\% missed in some cases
\citep{db82}.  HICA is thus a valuable probe of the cold \hi\ phase, but the
discrete nature of most continuum backgrounds compromises HICA's ability to map
the cold gas distribution.

A more suitable alternative is \hi\ {\it self-absorption}\footnote{In the
convention of 21cm line work, this long-used term (e.g., \citealt{w52,r60})
does not imply that the absorbing and emitting gas are intermixed or even close
together along the line of sight \citep*{blb78}, although these possibilities
are not excluded.}  (HISA), which arises from \hi\ gas with an excitation
temperature colder than the background \hi\ brightness temperature.  Since the
total Galactic \hi\ brightness is $\la 100-150$~K, HISA has a built-in
temperature filter that allows it to trace cold gas exclusively.  Like HICA,
HISA requires suitable backgrounds, but bright \hi\ emission is found over
large, continuous areas at low latitude.  As a result, HISA is an excellent
tool for mapping the structure and distribution of cold Galactic \hi.

\subsection{HISA Observations}

A thorough HISA survey of the cold \hi\ population requires high angular
resolution, high radial velocity resolution, and broad sky coverage.  The first
two provide a detailed view of HISA structure, with reduced dilution of
spatially and spectrally sharp absorption features, and refined estimation of
the unabsorbed \hi\ brightness, as measured near HISA feature edges in space
and velocity.  The third permits the study of a large population of objects,
unbiased by any target preselection (e.g., molecular clouds), allowing the cold
\hi\ properties to be examined in the context of the larger environment.

Due to the effort and resources required, a HISA survey incorporating all three
ingredients has never been published; the closest approximation is the Boston
University Arecibo \hi\ survey, only part of which is Nyquist sampled
(\citealt{kb94}; T. M. Bania 2005, in preparation).  Two prior Arecibo mapping
surveys consisted of longitude-velocity or latitude-velocity strip-maps only
\citep{bb79,bl84}.  Other HISA surveys have targeted known molecular clouds
with single-dish telescopes and little or no imaging
\citep*{k74,msb78,lb80,s81,stl88,lg03}, and until recently only a few isolated
objects have been observed with synthesis telescopes
\citep*{r72a,c75,lrh80,r80,r81,dr82,vdw88,v89,f93,s95,lr97}.

Fortunately, new technological advances have made \hi\ synthesis surveys over
large areas possible.  One such effort, the Canadian Galactic Plane Survey
(CGPS; \citealt{cgps}), is the basis of this paper.  The CGPS combines a
hexagonal grid of full-synthesis fields with single-dish observations to enable
the detection of all scales of \hi\ structure down to the synthesized beam.
The CGPS \hi\ data have a $58\arcsec \times 58\arcsec{\rm cosec}(\dec)$ beam,
0.824~\kms\ velocity sampling, and a field-center noise of $\trms \sim 3$~K in
empty channels; $\trms$ doubles when the 107\arcmin\ primary beam is filled
with 100~K emission, and it can be up to 60\% greater between field centers.
The initial phase of the survey mapped a $73\arcdeg \times 9\arcdeg$ region
along the Galactic plane with longitudes $74.2\arcdeg \le \glon \le
147.3\arcdeg$ and latitudes $-3.6\arcdeg \le \glat \le +5.6\arcdeg$
($+33.9\arcdeg \le \dec \le +68.4\arcdeg$), and extensions in both $\glon$ and
$\glat$ have followed.  This paper deals solely with the Phase~I CGPS, which we
shall call simply ``the CGPS'' for brevity.  Paper~I gave a preliminary view of
HISA in the CGPS, including many small-scale features masked by beam dilution
in single-dish surveys.  Larger-scale HISA features have also been found in
CGPS data by \citet{ht00}, \citet{kb01}, and \citet{h05}.  Two other major
synthesis surveys, the Southern Galactic Plane Survey (SGPS; \citealt{sgps})
and VLA Galactic Plane Survey (VGPS; \citealt{vgps}), have abundant and
intricate HISA populations as well \citep{mg01,k03,g04}.

\subsection{A Large-Scale HISA Survey}

Following the first look at CGPS HISA given in Paper~I, this paper presents a
large HISA survey over the entire $73\arcdeg \times 9\arcdeg$ CGPS.  Our intent
is to chart all of the cold \hi\ in the CGPS that is readily detectable as HISA
so that the feature population can be studied over a wide area with minimal
selection bias.

A companion paper (\citealt{g05a}; hereafter Paper~III) describes the automated
algorithms used in this paper to locate HISA features in the \hi\ data cubes
and extract them for analysis.  Our technique seeks out dark features in the
\hi\ data with abundant small-scale structure in space and velocity.  This
approach succeeds in finding most of the HISA that is conspicuous to the eye.
Maps of the detected HISA trace a rich array of cold \hi\ features over the
breadth of the CGPS, including a froth of low-amplitude HISA and a number of
elaborate complexes of stronger absorption.
The weak absorption seems well mixed with the general \hi\ emission and may
arise from small turbulent fluctuations in velocity and temperature in the ISM.
Many of the stronger absorption features appear to probe cold atomic gas in the
velocity reversal of the Perseus spiral arm and may trace evolving material in
the arm.

Below, we summarize our method of HISA identification and extraction
(\S\ref{Sec:ext}), present the features it finds in the CGPS
(\S\ref{Sec:maps}), and consider the spatial and velocity distributions of the
HISA (\S\ref{Sec:dist}).  Detailed analyses of the properties of the HISA
features and their degree of correspondence with molecular gas will be
presented separately \citep{g05b}.  Early results from some of this work have
been discussed in \citet{g02}.

\section{HISA Identification and Extraction}
\label{Sec:ext}

In past HISA studies, visual inspection sufficed for identifying
self-absorption features, but the large amount of data in a survey like the
CGPS makes this impractical.  Visually identified HISA is also unlikely to be
selected in a uniform or repeatable manner.  To address these concerns, we have
developed an automated means of searching for HISA in 21cm spectral line data.
We have also developed a method of interpolating the $\toff$ brightness around
the feature in $(\glon,\glat,\vel)$ space to obtain the best-constrained
estimate of the unabsorbed brightness $\tu$ that would be observed at the
feature position in the absence of HISA.  From this, the absorption amplitude
of the feature is obtained as $\dt \equiv \ton - \tu$, where $\ton$ is the
brightness temperature on the feature, and $\dt < 0$.  Full discussions of the
workings and evaluation of our HISA search and extraction software are
presented in Paper~III, but we summarize key aspects here.

Our search method is conservative: we seek \hi\ features that can only be
identified as HISA.  Following the criteria of Paper~I, such features have
narrower line widths than most emission, steeper line wings, more small-scale
angular structure, and a minimum background \hi\ brightness.  The last
condition excludes narrow gaps between sharp-edged emission features, which are
more common in the absence of bright backgrounds.  Using a variant of the CLEAN
algorithm of \citet*{sdi84}, we iteratively remove large-scale emission
features from the \hi\ data and flag significant negative residuals as HISA.
The CLEAN is performed separately in the spectral and spatial domains, and only
features satisfying both searches are retained.  $\tu$ is then estimated by
interpolating the $\toff$ brightness around the feature in 3-D.  HISA
detections with $\tu < 70$~K are rejected.  Detections where the survey noise
is unacceptably high are also rejected as untrustworthy.  The CLEAN filtering
removes HISA on the largest spectral and spatial scales, but some of this
larger structure is recovered by subtracting the flagged absorption from the
\hi\ data and rerunning the search.  A total of 3 search+extraction iterations
were applied to the CGPS data.

The search software was tuned to find as much visually identifiable HISA as
possible while minimizing false detections.  It was then tested extensively
with model \hi\ data containing random backgrounds, absorption features, and
noise.  The absorption extracted by the software was compared to that put into
the models, and the software performance was evaluated as a function of four
``observables'': $\dt$, $\tu$, $\da$, and $\dv$, where the latter two measure
the local angular and velocity widths of features.  Within this 4-D parameter
space, we measured $\fid$, the fraction of input HISA detected, $\fot$, the
fraction of output HISA that represents true detections, and $\ddt$, $\dda$,
and $\ddv$, the measurement drifts between input and output ($\dtu = -\ddt$).

We found that HISA meeting our design criteria is generally well detected:
$\fid$ is high for HISA significantly stronger than the noise level, larger
than a few beams, narrower than a few \kms, and with $\tu \ga 80$~K.  At the
same time, the bulk of HISA detections are reliable, with $\fot$ near unity in
most parts of the parameter space except those occupied by beam-scale noise
fluctuations.  Measurement drifts are small in well-detected features, with
$|\dt|$ and $\tu$ being underestimated by a few K due to contamination of
$\toff$ by faint, undetected HISA near the feature.  Where detections are
truncated by noise fluctuations or faint $\tu$, the bias may be somewhat
larger.  $\dda$ and $\ddv$ are both negative in most cases, since incomplete
detection makes features appear smaller and narrower.  In practice, these
effects manifest as the stronger-absorption cores of HISA features being
detected when some weaker-absorption envelopes may be missed.

\section{Absorption Amplitude Maps}
\label{Sec:maps}

\subsection{Constraints on the Survey Area}
\label{Sec:maps_overview}

\begin{figure}
\plotone{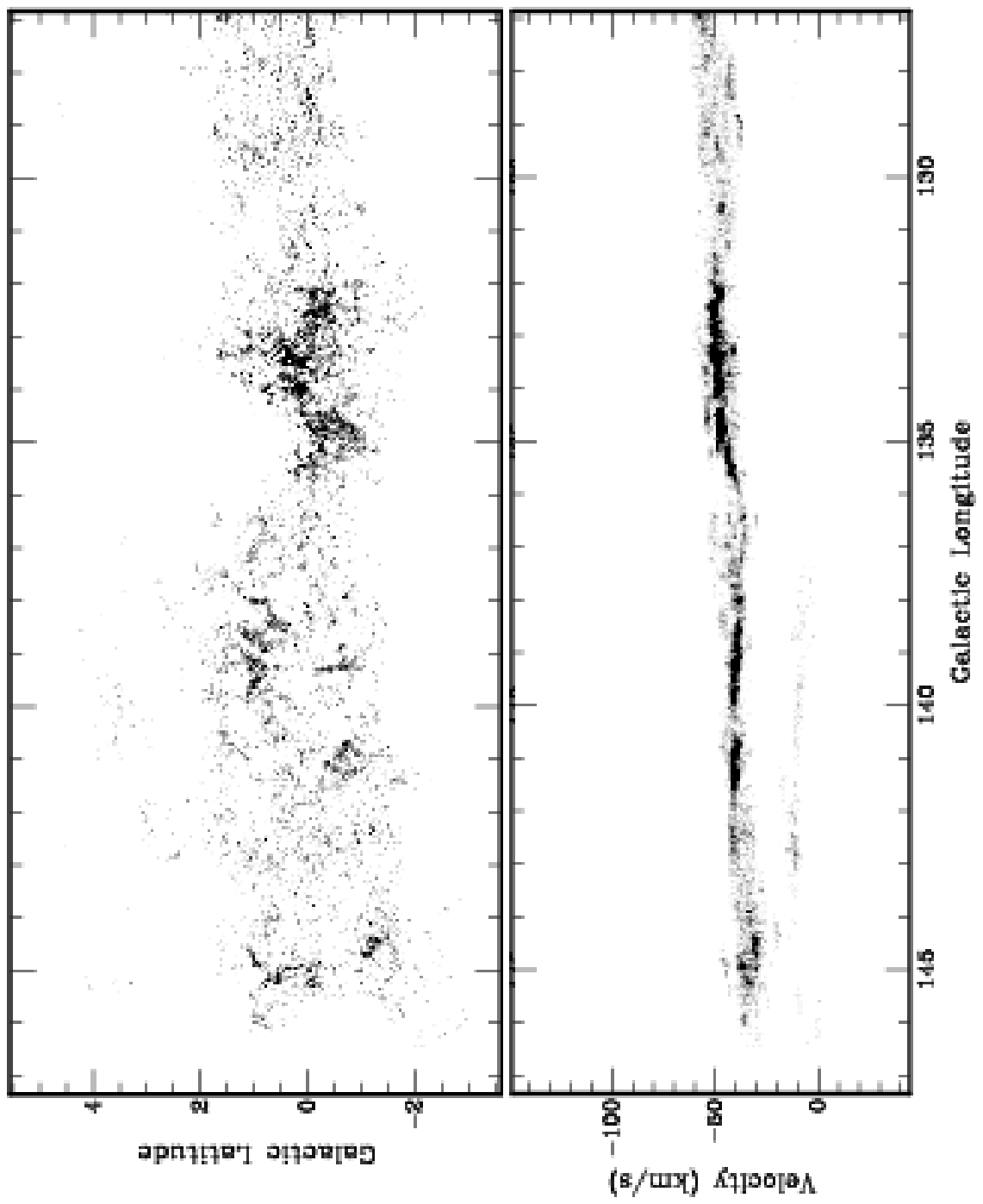}
\caption[]{
{\it (a)\/} 
Detected HISA in the CGPS over the longitude range 
$127\arcdeg \le \glon \le 147\arcdeg$, showing the amplitude $\dt$ in 
$(\glon,\glat)$ integrated over $\vel$ 
and in $(\glon,\vel)$ integrated over $\glat$.
Intensity ranges are $-82.446$ to 0 \kkms\ and $-5.0$ to 0 \kdeg, respectively.
{\it (b-d)\/} Same as {\it (a)}, except covering 
$109\arcdeg \le \glon \le 130\arcdeg$, 
$92\arcdeg \le \glon \le 112\arcdeg$, and 
$74\arcdeg \le \glon \le 94\arcdeg$, respectively.
The longitude ranges match those shown in Figure~7 of \citet{cgps}.
}
\label{Fig:hisa_overview_map}
\end{figure}

The CGPS \hi\ data are released in $5.12\arcdeg \times 5.12\arcdeg \times
211\kms$ ($1024 \times 1024 \times 256$) cubes.  All 36 of the Phase~I cubes
were processed with the HISA identification and extraction software described
in \S\ref{Sec:ext}.  The results were then merged into a single data set
covering the full CGPS\footnote{Electronic copies of these data are available
upon request.}.  Figure~\ref{Fig:hisa_overview_map} shows side-by-side
$(\glon,\glat)$ and $(\glon,\vel)$ projections of these data, in which $\dt$ is
integrated over $\vel$ and $\glat$ respectively.  The $\dt(\glon,\glat,\vel)$
data are sparse enough that most features are readily located in 3-D by
comparing the two projections.  Figure~\ref{Fig:hisa_overview_map} is divided
into four longitude subranges matching those of Figure~7 of \citet{cgps}.

\subsubsection{Noise Limits}
\label{Sec:maps_overview_area}

The total area mapped by the 193 overlapping fields of the CGPS is
656.2~deg$^2$, but this includes synthesis field peripheries with substantial
noise.  To maximize the reliability of our survey, HISA voxels were required to
have CGPS field mosaic weights $\wmos \ge 0.382$, the lowest weight that occurs
between synthesis field centers.  Since $\wmos \propto {\sigma_{noise}^{-2}}$,
this allows a maximum noise of 1.618 times the field center value (see
\citealt{cgps}), which is typically $5-7$~K for the $\tu \sim 70-130$~K levels
of our HISA features.  Such peripheral clipping and the additional exclusion of
20\arcmin\ strips at mosaic edges by the spatial search algorithm (see
Paper~III, Fig.~3) reduce the search area to 528.3 deg$^2$ in a region bounded
by $75.0\arcdeg \le \glon \le 146.5\arcdeg$ and $-3.0\arcdeg \le \glat \le
+5.0\arcdeg$, with complete coverage for $76.0\arcdeg \le \glon \le
145.5\arcdeg$ and $-2.5\arcdeg \le \glat \le +4.5\arcdeg$, except near the
bright radio source Cas~A $(\glon=111.7\arcdeg,\glat=-2.1\arcdeg)$.
Figure~\ref{Fig:survey_edges_and_c21} marks the boundaries of this area.

\subsubsection{HICA Contamination}
\label{Sec:maps_overview_cont}

Figure~\ref{Fig:survey_edges_and_c21} also plots contours of 21cm continuum
emission (shown in grayscale in Figure~7 of \citealt{cgps}).  Where the
continuum is bright, HICA may ``contaminate'' the cold \hi\ sampled by HISA in
two different ways: by adding absorption where the \hi\ background is too faint
for HISA, or by allowing warmer gas to contribute to the absorption than HISA
would by itself, since the maximum temperature of the absorbing gas is the
brightness temperature of the background \hi\ $+$ continuum.  Both problems are
exacerbated by the spotty nature of most bright continuum emission, which adds
artificial absorption structure to the HISA map.

Our HISA identification software excludes sight lines with HICA strong enough
for $\ton$ to be significantly less than zero (Paper~III).  Apart from special
cases involving polarization \citep*{klw04}, weaker HICA cannot be separated
cleanly from the HISA without knowing the properties of the absorbing gas.  In
the 4-component radiative transfer model of Paper~I, the total HISA+HICA
amplitude is

\begin{equation}
\label{Eqn:dt_4comp}
\dt_{tot} \;\; = \;\;
\left( \ts - \tc - \bfrac \, \tu \right) 
\left( 1 - e^{-\tau} \right)
\;\; ,
\end{equation}

\noindent where $\ts$ is the ``spin'' (excitation) temperature of the absorbing
gas, $\tau$ is its optical depth, $\bfrac$ is the fraction of $\tu$ originating
behind the absorbing gas, and $\tc$ is the continuum brightness temperature;
all of $\tc$ is assumed to originate in the far background for simplicity.  The
absorption from HISA alone is found by setting $\tc = 0$ in
Equation~(\ref{Eqn:dt_4comp}), so its fraction of the total is

\begin{equation}
\label{Eqn:hisa_fraction}
\frac{\dt\subhisa}{\dt_{tot}} \;\; = \;\;
\frac{\ts - \bfrac \, \tu}{\ts - \tc - \bfrac \, \tu}
\;\; .
\end{equation}

\noindent For $\toff = 100$~K, $\tc = 20$~K, and $\bfrac \le 1$, this ratio is
$\le 0.71$ for $\ts = 50$~K and $\le 0.82$ for $\ts = 10$~K.  Higher fractions
obtain if some of the continuum emission lies in front of the absorbing gas.
Constraining the radiative transfer adequately to separate HISA from HICA is
beyond the scope of this paper.  Fortunately, most detected CGPS HISA is
largely HICA-free, with only a smooth, faint $\tc \sim 5$~K background from
Galactic synchrotron and cosmic microwave background radiation.  The two most
significant exceptions are a small area around the bright \hii\ region W3
($\glon=133.81\arcdeg,\glat=+1.22\arcdeg$; e.g., \citealt{nor99}) and a larger
area in the extensive Cygnus~X complex of \hii\ regions and supernova remnants
($77\arcdeg \la \glon \la 83\arcdeg, -1\arcdeg \la \glat \la +3\arcdeg$; e.g.,
\citealt*{whl91}).  The bulk of the remaining HICA is in compact features
scattered across the CGPS area (Fig.~\ref{Fig:survey_edges_and_c21}; see
\citealt{st04} for a survey).  A prominent example is the ultracompact \hii\
region DR~7 ($\glon=79.32\arcdeg,\glat=+1.29\arcdeg$; \citealt{ww96}), which
produces HICA over a wider velocity range than the HISA near $\vlsr \sim
0~\kms$ in Figure~\ref{Fig:hisa_overview_map}.

\subsubsection{$\tu$ Restriction}
\label{Sec:maps_overview_tu}

In addition to the boundary marked in Figure~\ref{Fig:survey_edges_and_c21},
HISA detections are also restricted by the requirement that $\tu \ge 70$~K,
since HISA needs bright \hi\ backgrounds for it to arise and to be clearly
detected (\S\ref{Sec:ext}).  The effects of this constraint are illustrated in
Figure~\ref{Fig:hisa_vs_hie}, which plots approximate $\tu = 70$~K contours
over $(\glon,\glat)$ and $(\glon,\vel)$ maps of detected HISA.  These contours
are taken from $T_{max}$ maps of 20\arcmin-smoothed CGPS \hi\ emission, where
$T_{max}$ is the brightest value along the projected axis, thus indicating
positions where suitable HISA backgrounds may occur.  The actual $\tu$ used in
the HISA identification software is difficult to show in 2-D, but these
smoothed maps give a reasonable indication of where HISA detections are
possible.  Their 70~K levels correspond closely to a number of sharp edges of
the HISA distribution.  The cold \hi\ gas traced by the HISA may continue
beyond these apparent edges in many cases.  Given the large number of Galactic
\hi\ sight lines and velocities for which $\tu < 70$~K, it is important to
emphasize that the detected HISA population may represent only a small subset
of the total cold \hi\ population.  The $(\glon,\glat)$ area enclosed by this
$\tu$ contour is 398.0 deg$^2$.  In \S\ref{Sec:dist}, we will consider the
fraction of this area covered by detected HISA of different strengths, along
with other survey completeness issues.

\subsection{General Appearance}
\label{Sec:maps_general}

A detailed analysis of the general HISA distribution is given in
\S\ref{Sec:dist}, and a full discussion of HISA gas properties and
correspondences with molecular material will be presented in a separate paper.
However, a few points require initial comment.

Figure~\ref{Fig:hisa_overview_map} shows a broad swath of HISA over the entire
length of the survey.  The detection of this HISA immediately confirms that gas
radial velocities do not change monotonically with distance in the outer Galaxy
(e.g., from pure differential Galactic rotation), since any HISA requires
background \hi\ emission at the same radial velocity as the absorbing gas.  The
detection of so {\it much\/} HISA over such a large area indicates this
situation is widespread.

Low-amplitude HISA appears almost ubiquitous, arising nearly everywhere that
$\tu$ is bright enough for HISA detection (Fig.~\ref{Fig:hisa_vs_hie}).  By
contrast, darker HISA is more concentrated in both space and velocity,
collecting in complexes that often resemble molecular clouds in their
structure.  Quite strong absorption is found at velocities consistent with both
Local and Perseus arm gas, although at very different longitudes.  In this
paper, ``strong'' simply means large $|\dt|$, which could be caused by a
variety of factors in Equation~(\ref{Eqn:dt_4comp}).

The strongest HISA is at low velocities and low longitudes, where it coincides
with three large CO complexes described by \citet{dt85}: the nearby Cyg~OB7 and
Cyg~Rift complexes and the more distant Cyg~X complex that lies beyond the Rift
along the tangent of solar motion.  The Local arm ($\vlsr > -20\kms$) also has
HISA at higher longitudes, including the Local HISA Filament of Paper~I, but
this is generally much weaker.  Whether there is less cold \hi\ here is
unclear, since $\tu < 70$~K in many areas.

Strong HISA is also found in the Perseus arm ($-20\kms > \vlsr > -70\kms$),
mostly at high longitudes.  The Perseus HISA Complex of Paper~I is now revealed
as one of a chain of HISA complexes over a large longitude range.  Many sight
lines within these detected HISA structures have no corresponding CO detection
(Paper~I), but the HISA velocity distribution is close to that of
\citeauthor{r72b}' (\citeyear{r72b}) Perseus arm spiral shock \citep{g02}, a
point we shall revisit in \S\ref{Sec:dist}.

We find little HISA of any kind for $\vlsr < -70\kms$, and only then for $\glon
\la 100\arcdeg$.  This HISA is probably in the Outer arm, although the
Outer/Perseus boundary is difficult to determine for $90\arcdeg \la \glon \la
100\arcdeg$.  In this part of the Galaxy, HISA detections are limited by $\tu <
70$~K in most areas (Fig.~\ref{Fig:hisa_vs_hie}), and where $\tu > 70$~K, much
of the background \hi\ emission is too finely structured for our search to
detect large-scale HISA.  As a result, the large HISA shell of \citet{kb01} is
missed, and only a handful of small features are found.  More Outer-arm HISA
may remain undetected, and still more cold \hi\ may lack sufficient background
emission to self-absorb at all.

\subsection{Catalog of Major Features}
\label{Sec:maps_feature_catalog}

The high resolution and large area of the CGPS allow the thorough mapping of
known HISA features and the discovery of many new ones.  We define ``features''
as HISA voxel (volume-pixel) clusters in $(\glon,\glat,\vel)$ that may or may
not correspond to physical ``clouds''.  Cautious interpretation is required,
because shadow images do not give a complete picture of reality \citep{plato}.
Our HISA features may arise from discrete clouds of cold \hi, from cold \hi\
concentrations within \hh\ clouds, from portions of either where the background
\hi\ is bright enough for HISA to arise, or from sight-line superpositions of
any of these sharing the same radial velocity.

We have compiled a catalog of the most significant HISA features found by our
search algorithms in the CGPS.  Using the full-survey $\dt(\glon,\glat,\vel)$
data cube (\S\ref{Sec:maps_overview}), we identified contiguous groups of HISA
voxels as features, in the same fashion used for $\tu$ estimation.  To minimize
HICA contamination, all sight lines with $\tc > 20$~K were excluded.  This
process identifies thousands of features, most of which are small,
low-amplitude ``fluff''.  The larger, darker features have complex structure
that often intertwines with or surrounds neighboring features in 3-D.  Since
such cases suggest a physical relationship, we have simplified the catalog by
merging any two features together if the $(\glon,\glat,\vel)$ boxes containing
each feature overlap and both features satisfy a minimum size requirement.
Each feature to be merged must have a HISA $(\glon,\glat,\vel)$ volume of $\ge
0.1\lbvunit$ $ = 4851$ CGPS voxels.  This restriction keeps the widespread
fluff from joining all of the detected HISA into a few unrealistically large
features that, e.g., include HISA in both the Local and Perseus arms.

The most significant features in this merged catalog, defined as those for
which the HISA volume is also $\ge 0.1 \lbvunit$, are marked in
Figure~\ref{Fig:major_features}.  There are 69 such features.  If we apply the
\S\ref{Sec:maps_general} criteria for spiral arm membership, 11 features are
Local arm, 50 are Perseus arm, and 8 are Outer arm members.  All 69 contain at
least some high-$|\dt|$ voxels, but we do not require feature voxels to have a
minimum $|\dt|$, since this would identify only feature ``cores'' rather than
whole features.  $\dt$ varies across even the most obvious and coherent
features, and strong HISA voxels are almost always enveloped in weaker HISA,
especially near feature $(\glon,\glat,\vel)$ edges.

Table~\ref{Tab:major_features} lists these Galactic HISA (GHISA) features with
measurements of position, extent, $(\glon,\glat,\vel)$ volume, $(\glon,\glat)$
area, 3-D integrated absorption, mean and maximum absorption, mean $\tu$, and
three ``diagnostics'' of completeness and reliability described in
\S\ref{Sec:maps_rel}.  Derived properties like $\ts$ and $\coldens$ will be
considered in a separate paper.  The features do not fill their extents
solidly; volume filling factors, defined as the volume of HISA voxels divided
by the product of the $\glon$, $\glat$, and $\vel$ extents, are $\sim 3-30\%$,
with the larger features having lower filling factors on average.  This
clumpiness is also reflected in the {\it contiguous\/} extents, which although
not tabulated are $\sim 2\arcmin - 5\arcmin$ and $\sim 2-3\kms$ on average, and
$\la 12\arcmin - 30\arcmin$ and $\la 5-8\kms$ in all but the largest 3 or 4
features.  The total HISA volume detected in the CGPS is $214.98\lbvunit$, of
which 54.5\% is in the features of Table~\ref{Tab:major_features}.  The total
CGPS HISA integrated absorption is $-2791.4\absintunit$, of which the listed
features account for 66.2\%.

A comparison against past HISA studies in the CGPS region shows that ours
reproduces most prior detections while yielding an abundance of new features.
Table~\ref{Tab:major_features} notes past observations of each feature, and
Table~\ref{Tab:previous_detections} summarizes previous HISA work in the CGPS
region, including the numbers of past features we have found.  Of those that we
miss, only two from \citet{r80} have $\tu > 70$~K; these have $\tu \sim 85$~K
instead, where model tests in Paper~III show that our detection sensitivity is
high but not 100\%.  More striking is that the majority of CGPS features in
Table~\ref{Tab:major_features} have {\it no\/} prior detections in the
literature.  Although the five most conspicuous features were previously known,
most features with volumes of $\la 1\lbvunit$ and absolute integrated
absorptions of $\la 10\absintunit$ are new detections.  Altogether, only 11 of
the features in Table~\ref{Tab:major_features} were detected before the CGPS,
and only 7 were even partially mapped.  None has been mapped as thoroughly as
in the CGPS.

A few caveats on the interpretation of these features should be emphasized.
First, as noted above, velocity crowding may superpose unrelated absorption
along the line of sight, especially near the Local-arm tangent in Cygnus
($75\arcdeg \la \glon \la 90\arcdeg, |\vlsr| \la 20\kms$).  Second, despite the
merging of some proximate features, $\tu$ variations may artificially fragment
large clouds into multiple features or limit their apparent extent, and the
survey edges themselves may truncate some features (e.g., GHISA
079.88$+$0.62$+$02, GHISA 091.90$+$3.27$-$03, GHISA 145.01$+$0.46$-$35, or
GHISA 145.81$+$0.96$-$36).  Finally, the CGPS HISA has a very complex
structure, and feature definitions under such circumstances are somewhat
subjective.  In particular, the choice of merging criteria affects the size and
number of major features identified, since if smaller features are allowed to
merge, fewer and larger features result.  We sought criteria that mimic the
eye's identification of features.  These criteria were applied algorithmically
to ensure precise measurements and uniform treatment.

\subsection{Reliability and Completeness}
\label{Sec:maps_rel}

The software performance measures $\fid$ and $\fot$ can be used to assess the
reliability and completeness of particular HISA feature detections.  For
reliability, the observables $\dt$, $\tu$, $\da$, and $\dv$ are measured at
each HISA voxel $(\glon,\glat,\vel)$ position, and the $\fot$ value
corresponding to these coordinates in the 4-D parameter space is assigned to
$\fot(\glon,\glat,\vel)$.  For each feature in Table~\ref{Tab:major_features},
we determined the fractions of its HISA volume and 3-D integrated absorption
that are reliably detected as

\begin{equation}
\label{Eqn:fotfracvol}
\fotfracvol \;\;\; = \;\;\;
\frac{1}{\nvox}
\, {\displaystyle \sum_{n=1}^\nvox {\fot(\glon_n,\glat_n,\vel_n)}}
\end{equation}

\noindent and

\begin{equation}
\label{Eqn:fotfracint}
\fotfracint \;\;\; = \;\;\;
\frac{\displaystyle \sum_{n=1}^\nvox
	{\fot(\glon_n,\glat_n,\vel_n) \cdot \dt(\glon_n,\glat_n,\vel_n)}}
     {\displaystyle \sum_{n=1}^\nvox{\dt(\glon_n,\glat_n,\vel_n)}}
\;\; ,
\end{equation}

\noindent respectively.  For the features listed, $0.64 \le \fotfracvol \le
0.98$ and $0.68 \le \fotfracint \le 0.98$, with respective medians of 0.86 and
0.90.  For the survey as a whole, $\fotfracvol = 0.80$ and $\fotfracint =
0.86$.  Since $\fot$ increases with $|\dt|$, $\fotfracint \ge \fotfracvol$
generally, and features with stronger $\dtavg$ typically have greater
reliability fractions.

A similar analysis of detection completeness in terms of $\fid$ is not
possible, because convergent drifts (e.g., $\dda \propto - \da$) prevent the
recovery of the HISA input properties $(\dti,\tui,\dai,\dvi)$ upon which $\fid$
depends.  However, completeness can be partially assessed in terms of the
measured property $\tuo$; for features of appreciable size and strength, $\dtu$
converges symmetrically about $\tu = 80$~K (Paper~III), and thus $\tuo \sim
\tui$ near 80~K.  Table~\ref{Tab:major_features} lists $\tuavg$, the average
$\tuo$ per voxel, and $\tufrac$, the fraction of voxels for which $\tuo \ge
80$~K.  $\fid > 0.8$ for reasonably strong features with $\tui > 80$~K but
declines rapidly for lower $\tui$ (Paper~III).  Although the amount of
undetected HISA cannot be quantified, $1 - \tufrac$ gives the fraction of each
feature lying where significant HISA might be missed.  For example, GHISA
139.01$+$0.96$-$40 (the Perseus HISA Complex of Paper~I) offers few
opportunities to hide strong, undetected HISA.  By contrast, GHISA
091.90$+$3.27$-$03 has a much lower $\tufrac$, and its high-latitude edge is
definitely truncated by $\tu < 70$~K.  For the entire CGPS HISA set, $\tuavg =
92.29$~K and $\tufrac = 0.75$.

\subsection{A Remarkably Large, Strong Feature}
\label{Sec:maps_feature_comments}

GHISA 091.90$+$3.27$-$03 is one of the most striking HISA features in the CGPS,
as shown in Figures~\ref{Fig:hisa_overview_map}, \ref{Fig:major_features}, and
\ref{Fig:real_hisa_id_stage_maps}.  It has the strongest maximum absorption
($-80.72$~K) in the survey and the strongest average absorption ($-24.02$~K) of
any major feature.  It also has the largest area of apparent ``solidity'', with
contiguous HISA extents of $\la 2\arcdeg$ in $\glon$ and $\glat$.  This dark
center covers a few square degrees, but it is surrounded by subtle knots and
filaments over much larger area, and the feature can even be traced beyond the
CGPS boundaries to $\glat \sim +6\arcdeg$ in the single-dish surveys of
\citet{hb97} and \citet{ht00}.  As indicated in Tables~\ref{Tab:major_features}
\& \ref{Tab:previous_detections}, GHISA 091.90$+$3.27$-$03 has been studied by
several groups and mapped with 25m-class telescopes.  It coincides with the
dust cloud complex Kh~141 \citep{k60} and many \citet{l62} dark nebulae,
including LDNe 999, 1002, 1005, and 1029.  It may also be associated with other
dust clouds over a wider area, since it appears to be part of the large Cyg OB7
giant molecular cloud complex mapped in \twco\ by \citet{dt85}.  Its HISA-CO
correspondence is illustrated in Figures~\ref{Fig:real_hisa_id_stage_maps} and
\ref{Fig:real_hisa_id_stage_spectra}.  This feature may be related to the even
larger, but more diffuse, GHISA 079.88$+$0.62$+$02 complex to the west, which
also has significant dust content (e.g., LDN 906).

GHISA 091.90$+$3.27$-$03 appears dark, cold and quiescent.  In many areas it
has $\ton \sim 30-40$~K, $\dt \sim -50$~K, and a line full width at half
maximum (FWHM) of $\dv \sim 2.5$~\kms; a few of the darkest HICA-free knots
have $\ton \sim 20-30$~K and $|\dt| \sim 60-70$~K.  The small white $6\arcmin
\times 6\arcmin$ box in Figure~\ref{Fig:real_hisa_id_stage_maps} marks an area
of fairly typical absorption.
For the 4-component radiative transfer geometry of Paper~I, the box average
values of $\tc = 7.7$~K, $\ton = 45.8$~K, and $\tu = 92.6$~K imply $\ts \le
53.5$~K for $\bfrac \le 1$ and $\tau \le \infty$.  Since the HISA line does 
not appear strongly saturated, more reasonable upper limits are probably 
$\ts \le 46.2 - 51.1$~K for $\tau \le 2 - 3$.  If $\bfrac \sim 1$, as seems 
likely for such a conspicuous feature, lower limits of $\ts \ge 2.73$~K or 
10~K imply $\tau \ge 0.65$ or 0.75, respectively.  These four $(\ts,\tau)$ 
combinations yield $N_{H I} \sim$ 0.086, 0.35, 4.4, and 
$7.4 \times 10^{20} \; \perarea$.
Previous authors have found $N_{H I} = 1.53$ \citep{k74}, 0.8 \citep{s76}, and
0.1 to $2.0 \times 10^{20} \; \perarea$ \citep{sot81} in different parts of the
feature.  However, the gas mass is probably dominated by \hh.  For a mean
molecular weight per \hh\ molecule of $2.76\,m_{\rm H}$, \citeauthor{dt85}'s
(\citeyear{dt85}) results imply an average molecular column of $N_{H_2} \sim
\scinot{2.3}{21} \; \perarea$.
Our measured HISA column represents only $\sim 0.19-14$\% of the mass in the
total gas column.
\citet{d87} estimated the total Cyg OB7 mass as $\sim \scinot{7.5}{5} \; \msun$
for a distance of $\sim 800$~pc.

\section{HISA Distribution Analysis}
\label{Sec:dist}

We now consider general trends in the distribution of HISA in the CGPS and
their significance.  Of particular interest is the relationship between HISA
and Galactic structure.  We begin with a quantitative discussion of how common
HISA is in our survey and what implications this may have for the general
picture of the interstellar medium.

\subsection{Covering and Filling Fractions}
\label{Sec:dist_coverage}

HISA is found in 3.8\% of the CGPS voxels that have $\tu > 70$~K, $\tc \le
20$~K, and adequate mosaic weighting as defined in
\S\ref{Sec:maps_overview_area}.  21\% of the $(\glon,\glat)$ sight lines 
passing through this searchable volume have detected HISA, and 47\% of the
$(\glon,\vel)$ positions have detected HISA.  
A relatively small part of the CGPS volume is searchable for HISA, because
$\tu$ is high only in particular regions.  However, this portion is still large
enough to explore significant characteristics of the HISA distribution in the
outer Galaxy.  Detection reliability does not significantly affect these
results, since $\fotfracvol = 0.80$ for the full survey (\S\ref{Sec:maps_rel}).

A 3.8\% occupation of the $(\glon,\glat,\vel)$ volume may seem low, but in fact
this is quite a large fraction: it implies a linear filling factor of
$0.038^{1/3} \sim 0.33$ in each dimension for a uniformly random distribution.
These values are almost certainly lower limits, since $\fid \ll 1$ for $|\dt|
\ll \trms$, and the abundance of HISA with $|\dt| \sim \trms$ (Paper~III,
Fig.~11) suggests that fainter HISA may be at least as common as
HISA at the noise level.

Figure~\ref{Fig:hisa_coverage} plots the relative incidence of all CGPS HISA
voxels with $|\dt|$ above a given threshhold.  Curves are given for all voxels
and for those with $\fot \ge 0.90$ and 0.99.  The latter two respectively
eliminate about one-half and about two-thirds of the total voxel count, mostly
at the weak $|\dt|$ end where noise contamination is significant.  These curves
may be rescaled by the occupation fractions listed above to find the fraction
of the survey that contains HISA of a particular minimum strength.  For
example, the volume occupied by reasonably strong HISA with $|\dt| > 25$~K is
$\sim 10\%$ that of the total HISA population, while HISA with $|\dt| > 50$~K
accounts for less than 1\% of the total.

Simplisticly, we might assume that the HISA census is complete above some
$|\dt|$ level and extrapolate linearly in log space to lower $|\dt|$ to find
the amplitude for which the volume filling reaches 100\%.  E.g., for
$|\dt|_{complete} \sim 10$~K, we find $|\dt|_{filled} \sim 0.1$~K.  This
extrapolation is not very trustworthy, since it neglects $\ddt$ and the more
complicated dependences of $\fid$, but it suggests that at low enough
(sub-Kelvin) amplitudes, HISA may be truly ubiquitous.  Previously,
\citet{dl90} proposed that faint HISA may cause much of the structure in
low-latitude Galactic emission profiles, and \citet{d03} have argued that HISA
occurs in most \hi\ spectra, even if it is not always detectable.  Our CGPS
HISA census is consistent with these views.  At the very least, our survey
approaches the confusion limit for HISA detections.

\subsection{Weak and Strong HISA}
\label{Sec:dist_weakstrong}

As noted above, HISA with low $|\dt|$ is more frequent and more smoothly
distributed, while stronger HISA is more rare and more clumped into distinct
features.  This behavior can be inferred from the $\dt$ maps in
Figure~\ref{Fig:hisa_overview_map}, the incidence curves in
Figure~\ref{Fig:hisa_coverage}, and the loose $(\dt,\da)$ and $(\dt,\dv)$
property correlations in Paper~III, Figure~11.

Figure~\ref{Fig:strength_vs_dispersion} shows the relationship between HISA
amplitude and scale more explicitly.  Here we plot the RMS dispersions
$\sigma_\rho$ and $\sigma_\vel$ of the angular and velocity separations between
voxel pairs in the survey.  For every HISA voxel in every CGPS mosaic, the
distances to its HISA neighbors along lines of constant $\glon$, $\glat$, and
$\vel$ were measured, and RMS variations of these measurements were binned by
$|\dt|_{min}$, the absolute amplitude of the fainter of the two voxels in
question.  The results, combined for all 36 mosaics, show that weak HISA is
more diffusely distributed than strong HISA in both space and velocity.  This
is not a contradiction of the loose correlation of $|\dt|$ with $\da$ and $\dv$
in Paper~III, Figure~11, since those measured {\it contiguous\/} rather than
general extent.  Instead, the two trends together imply that strong HISA is
concentrated into more solid complexes, while weak HISA is spread out in a more
porous fashion.

The maximum dispersion scales for the weakest HISA are $\sim 2\arcdeg$ and
$\sim 5\kms$.  Once $|\dt|_{min}$ exceeds the \hi\ noise level of $\sim 6$~K,
both dispersions drop steadily, with the steepest descent near $\sim 20$~K.  At
higher amplitudes, the descent slows and approaches pseudo-asymptotes of $\sim
0.75\arcdeg$ and $\sim 0.8\kms$
The former is a typical scale for strong HISA features, while the latter
corresponds to a 1-channel separation.  Above $\sim 65$~K, angular coherence is
lost, the features break into smaller pieces that quickly approach the beam
scale, and even 1-channel separations start dropping out when only the maximum
absorption channel is sampled.  When converted to Gaussian FWHM, the maximum
velocity scale approaches the velocity width of the brightest part of a spiral
arm in \hi\ emission, and the minimum velocity scale approaches the narrowest
linewidths that can be sampled by the CGPS.  Angular FWHM are less physically
meaningful since the HISA angular structure is complex, but the maximum angular
scale is consistent with the $\sim 5$\arcdeg\ size of the CGPS mosaics in which
the measurements were made, implying that the weakest HISA has no preferred
angular distribution either.  Thus, the general trend in both angle and
velocity is from weak HISA ``fluff'' (\S\ref{Sec:maps_feature_catalog})
dispersed over the entire measurable area down to discrete, strong HISA
features.

The curves in Figure~\ref{Fig:strength_vs_dispersion} include the entire CGPS
HISA data set, but similar trends are seen in different spiral arms when
analyzed separately, or at different longitudes in different mosaics.  Only the
range of strongest absorption and the scale of maximum dispersion change from
one location to another.  For example, the Local-arm dispersions are larger in
both angle and velocity than the Perseus-arm dispersions, and more strong HISA
is found in the Local arm, at least in GHISA 079.88$+$0.62$+$02 and GHISA
091.90$+$3.27$-$03.

\subsection{Probing the Gas Kinematics}
\label{Sec:dist_kinmech}

HISA offers a unique opportunity to study gas motions in the outer Galaxy,
because its radiative transfer requires background emission at the same radial
velocity as the absorbing \hi.  HISA detections with $90\arcdeg < \glon <
270\arcdeg$ require departures from pure circular differential rotation, which
predicts only one distance per $\vlsr$ in this longitude range, apart from the
degeneracy at $\glon = 180\arcdeg$.  Two types of CGPS HISA are found: (1)
strong HISA with its accompanying weak $(\glon,\glat,\vel)$ envelopes, and (2)
unassociated ``fluff'' that makes up the rest of the weak HISA population.  We
suggest that these two types are produced by different mechanisms.

\subsubsection{Turbulence}
\label{Sec:dist_kinmech_turb}

Turbulence adds small-scale, random perturbations to the Galactic velocity
field.  These are likely to be very common but should have small velocity
amplitudes and little spatial coherence.  Turbulence may also produce cold gas
for absorption via departures from equilibrium conditions (e.g.,
\citealt*{svg02}; \citealt{j04}).  We propose that turbulent $(T,\vel)$
fluctuations are the source of the weak HISA ``fluff'' in the CGPS.  The fluff
lacks coherent large-scale structure, and it occurs in all parts of the CGPS
where $\tu > 70$~K, suggesting that it is ubiquitous and well-mixed with the
general \hi.

Turbulence cannot explain the strong HISA and weak envelopes; this absorption
is concentrated into discrete and often large-scale features with well-defined
edges.  Although brighter $\tu$ makes it easier to see, such HISA does not
appear everywhere that $\tu$ is bright enough, so a more organized process than
turbulence is needed.  The large amplitude of strong HISA is also problematic
for the turbulent explanation.  High $|\dt|$ can result from a high $\tc$,
$\tu$, $\tau$, or $\bfrac$ or a low $\ts$ (Eq.~\ref{Eqn:dt_4comp}).  High $\tc$
is rare in CGPS HISA, and we have excluded it where it might affect our
analyses.  $\dt$ and $\tu$ are poorly correlated at best (Paper~III, Fig.~11),
and neither $\tu$ nor $\bfrac$ is likely to change abruptly and frequently
enough to produce the sharp edges of strong HISA features.  Thus strong,
discrete HISA must arise from significant changes in $\tau$, $\ts$, or both.
In practice, these absorption enhancements resemble clouds or cloud complexes.

\subsubsection{Spiral Density Waves}
\label{Sec:dist_kinmech_wave}

Spiral density waves add large-scale, organized perturbations to the Galactic
velocity field.  These are less common than turbulent fluctuations but should
have large velocity amplitudes and significant spatial coherence.  They may
also be a natural source of cold gas.  Spiral waves and their associated shocks
have long been thought to drive the formation of dense interstellar clouds and
massive stars in galactic disks \citep{ls64,r69,s72}, but the full details of
the gas dynamics and evolution are complex and difficult to study.  In face-on
external galaxies, spiral structure is simple to observe, but spatial
resolution is limited by distance (e.g., M81: \citealt{aw96}).  By contrast,
spatial resolution is much better in the Milky Way, but our internal
perspective makes the line of sight geometry difficult to untangle.  HISA's
sensitivity to velocity reversals is an asset in this situation.

The nearby Perseus arm gives a detailed view of spiral structure and the
density wave phenomenon.  \citet{r72b} modeled the gas distribution and
kinematics in the Perseus spiral shock.  Figure~\ref{Fig:roberts_shock}
illustrates his model's gas density and velocity vs.\ distance along the line
of sight and the $(\glon,\vel)$ distribution of the gas passing through the
arm.  HICA studies have presented evidence in support of this model in a few
sight lines toward bright continuum sources \citep{g73, gl79, fh91, nor99,
kk02, b02}.  A more complete picture of the velocity reversal is possible with
HISA, which has a much more extensive background than HICA.  In the CGPS,
strong Perseus HISA is visible over a wide area (e.g.,
Fig.~\ref{Fig:hisa_overview_map}), as would be expected for a neighboring
spiral arm.

Figure~\ref{Fig:hisa_lv_shock} compares the $(\glon,\vel)$ distribution of
strong ($|\dt| > 20$~K) HISA to the spiral ``shock ridge'' of \citet{r72b}.  In
his model, the gas slows as it enters the arm shock, reaches a maximum
negative-velocity departure from the standard rotation curve speed, and then
rebounds toward a positive-going ``velocity hill'' and the normal rotation
speed as it moves on through the rest of the arm and out the back
(Fig.~\ref{Fig:roberts_shock}).  We propose that the strong HISA is in or
slightly downstream of the shock ridge, the point just behind the shock where
the gas is densest, and that \hi\ emission on the far side of the velocity hill
serves as a background with the same velocity as the absorbing gas.
In support of this assertion, most of the strong Perseus HISA lies on the shock
ridge or at more positive velocities for $110\arcdeg \la \glon \la 145\arcdeg$
(Fig.~\ref{Fig:hisa_lv_shock}).  Considering that the shock model does not
address structural inhomogeneities in the ISM
(e.g., \citealt{r93} and references therein),
the agreement is quite reasonable; even without prior lumpiness, the shock
process itself is expected to introduce some $(\glon,\vel)$ structure in the
cold gas distribution \citep{ki02}.

The Local arm is a more ambiguous case.  Its two largest features, GHISA
079.88$+$0.62$+$02 and GHISA 091.90$+$3.27$-$03, are associated with known CO
complexes either near the Sun or along the Local arm tangent.  The local CO
clouds are identified with the Great Rift system of dust clouds \citep{dt85},
which may be related to a ring of expanding material in Gould's Belt
\citep{l67,l73}.  Alternatively, GHISA 091.90$+$3.27$-$03 has been suggested to
trace a spiral shock in the Local arm \citep{s76}.  This scenario conflicts
with that of a Local arm shock in the Galactic center direction \citep{qc73},
but it is consistent with stellar streaming results \citep{sm99} and naturally
explains the presence of background \hi\ emission with the same velocity.
However, the Local arm is probably not a major spiral feature, and the dynamics
of its stars and gas may instead have been shaped by passage through a ``real''
spiral arm 100~Myr ago \citep{o01}.  

Whichever picture is correct, the presence of background emission at the right
velocites must be explained.  While the tangent of Solar motion enhances HISA
visibility in some areas, it is not clear that the tangent can provide adequate
backgrounds for all of the strong Local HISA; e.g., most of GHISA
091.90$+$3.27$-$03 has $\glon > 90\arcdeg$.  A Local spiral density wave shock
might provide backgrounds at such longitudes, but no quantitative model
predictions are available to compare against the CGPS HISA data.

As a general caveat, we note that one well-organized shock per arm may be
simplistic: non-grand-design galaxies may have more flocculent shock structure,
as suggested by the multiplicity of dust lanes in {\it Hubble Space Telescope}
images of the backlit galaxy NGC 3314 \citep{kw01}, and even some presumably
grand-design spirals may be more complicated (e.g., M51: \citealt{h03}).  The
Milky Way is unlikely to be a grand-design spiral, and arguments for multiple
shocks in the Perseus arm \hi\ data have been made \citep{b02}.

HISA features in the inner Galaxy have also been proposed to trace spiral shock
material \citep{s92,m01,mg01}.  More recently, \citet{g04} reported HISA
tracing what appears to be several spiral arms over a large longitude range in
the inner Galaxy.  These results will be presented in more detail in a separate
paper on HISA mapped with the VLA Galactic Plane Survey (S. J. Gibson et al.,
in preparation).

\subsection{An Evolutionary Scenario}
\label{Sec:dist_evol}

Strong HISA requires high $\tau$, low $\ts$, or both
(\S\ref{Sec:dist_kinmech_turb}).  The strong Perseus HISA $(\glon,\vel)$
distribution appears to trace cold \hi\ immediately downstream of the spiral
shock described by \citet{r72b}.  Passage through a spiral density wave could
cause a drop in gas temperature and an increase in density relative to the
ambient diffuse atomic medium.

In the classical picture, neutral atomic gas enters the trailing edge of a
spiral arm and is decelerated and compressed in the spiral shock; its increased
density then leads to rapid cooling, \hh\ condensation (e.g.,
\citealt{ki00,b04}), and star formation.  Elements of this process have been
studied by \citet{ht98} in CO and \hi\ emission in the W3/4/5 region of the
Perseus arm, and HISA in this area has been proposed to trace sites of
molecular formation \citep{hsf83,s90}.  The CGPS allows this phenomenon to be
investigated over a much wider area.  Since the Perseus arm HISA is at the
right velocity to trace the cold \hi\ that may be forming \hh, and it often
lacks CO detections at the same positions \citep{g00,g02}, it may probe the
atomic-to-molecular phase transition in the spiral shock environment, or at
least \hh\ clouds that have not yet formed CO, which could take much longer
\citep{b04}.

Alternatively, density waves may only concentrate pre-existing molecular gas
rather than causing an actual phase transition \citep*{vks88,pal01}.  In this
picture, HISA does not arise from ``pre-molecular'' clouds, but instead from
purely atomic cold \hi\ clouds or trace \hi\ in preexisting \hh\ clouds that
have been ``piled up'' by the spiral density wave without undergoing a phase
change.  \citet{ta91} argued that this is occurring in M51, where the peak \hi\
brightness in the arms falls slightly ``downstream'' of the densest regions
traced by CO and nonthermal radio continuum emission.  However, they noted that
their interpretation could be altered if the densest regions contained cold,
finely-structured \hi, which would not dominate the brightness temperature.
Our CGPS HISA results indicate that this is quite plausible.
\citeauthor{ta91}'s (\citeyear{ta91}) \hi\ observations were not sensitive to
spin temperatures of $14-45$~K, which are similar to those found for strong
Perseus HISA in Paper~I, and their physical resolution of $590 \times
820$~pc$^2$ in M51 would cover a $14 \times 19\,\ddeg$ area of the Perseus arm
at a distance of 2.5~kpc; this is several times larger than the $\sim
60\,\ddeg$ integrated $(\glon,\glat)$ area of the CGPS Perseus HISA.  Still,
even if considerable cold \hi\ is hidden in the spiral shock regions of M51, it
is a separate question whether this \hi\ is in mostly atomic or mostly
molecular clouds; if the latter, the phase change picture remains open to
question.

We will explore these issues in greater detail in a separate paper
investigating the gas properties and degree of molecular association of the
CGPS HISA \citep{g05b}.

\section{Conclusions}
\label{Sec:conclusions}

In this paper, we have presented a comprehensive survey of \hi\ self-absorption
(HISA) features in the initial $73\arcdeg \times 9\arcdeg$ phase of the
arcminute-resolution Canadian Galactic Plane Survey (CGPS).  The CGPS HISA is a
new tool for examining the detailed structure and distribution of cold \hi\
clouds in a large area of the outer Galactic disk.

Using CLEAN-based spatial and spectral filter algorithms and full 3-D
interpolation of the OFF-feature brightness to estimate the unabsorbed
intensity along the feature sight line, we extracted a set of HISA features
that extend over the entire area of the survey where conditions allow
detection, with sensitivity limits consistent with the noise and resolution
properties of the \hi\ data.  

We compiled a catalog of contiguous, isolated features exceeding a minimum
total absorption threshhold.  A comparison of this catalog against prior HISA
studies within the CGPS borders shows that we reproduce the majority of prior
discoveries, with misses generally due to faint backgrounds occurring outside
the range of our defined search parameters.  At the same time, we identify a
large number of new features not previously reported and image all HISA
features in unprecedented detail.  Small HISA structures can now be examined in
the context of larger ones, and many trends in the general distribution become
clear, such as the the concentration of strong HISA into a couple of very large
complexes at Local velocities, and a similar concentration of strong HISA into
a long series of complexes in the Perseus arm.

Statistically, the HISA detections fill a significant fraction of the
searchable volume and suggest that the faintest HISA may be ubiquitous in the
Galactic plane.  This weak HISA ``fluff'' also appears to have little
large-scale coherent structure in space or velocity but is as widespread as the
available backgrounds allow.  By contrast, strong HISA, which is much less
abundant, is concentrated into larger, more discrete, and relatively solid
structures.

Although the weak fluff and strong features occur within a continuum of
amplitudes and other properties, we suggest they are made visible by two
separate mechanisms.  Physically, the weak fluff may be interpreted as a
low-intensity population of small, slightly cool features scattered throughout
the ambient ISM, which are produced and revealed by turbulent fluctuations in
temperature and velocity.  The strong HISA, on the other hand, requires
significant density enhancements and/or very cold gas to produce its large
absorption amplitudes, and it needs a large-scale background at the right
velocity to absorb against in the outer Galaxy.  We propose that the velocity
reversal of the Perseus arm's spiral density wave provides the background for
the strong HISA in this location, while the strength of the absorption is due
to compression and cooling from the spiral shock.  Spiral arms have long been
thought to host the transformation of atomic gas into molecular clouds prior to
forming new stars, but this has been very difficult to study in our own Galaxy.
If HISA is revealing cold \hi\ evolving into \hh\ in a nearby spiral arm, this
offers the exciting prospect of examining the molecular condensation process in
great detail.

This paper is the second in an ongoing series investigating HISA at high
resolution in the Galactic plane.  A companion paper (Paper~III) describes the
HISA search and extraction algorithms at length, along with detailed
evaluations of their reliability and completeness.  A subsequent paper will
analyze the CGPS HISA and CO distributions together and the gas properties
implied for the HISA by available constraints.  A separate survey of the VGPS
HISA in the inner Galaxy is also in preparation, as are extensions to the CGPS
HISA survey coverage with new \hi\ observations.

\acknowledgements

We thank W. McCutcheon, T. Landecker, and J. Stil for a number of useful
discussions on this project, and the anonymous referee for constructive
comments on the manuscript.  W. Roberts graciously allowed the reproduction of
the plots in Figure~\ref{Fig:roberts_shock}, T. Dame generously provided the CO
survey data used in Figures~\ref{Fig:real_hisa_id_stage_maps} \&
\ref{Fig:real_hisa_id_stage_spectra} , and S. Strasser kindly made available
the 21cm continuum maps used in Figure~\ref{Fig:survey_edges_and_c21}.  We are
very grateful to R. Gooch for tireless computing support, including continued
expansion of the capabilities of the Karma visualization software package
\citep{karma96}\footnote{See also \url{http://www.atnf.csiro.au/karma}.}, which
was used extensively for this work.  The Dominion Radio Astrophysical
Observatory is operated as a national facility by the National Research Council
of Canada.  The Canadian Galactic Plane Survey (CGPS) is a Canadian project
with international partners.  The CGPS is described in
\citet{cgps}\footnote{Additional information is available online at
\url{http://www.ras.ucalgary.ca/CGPS}.}.  The main CGPS data set is available
at the Canadian Astronomy Data Centre\footnote{See
\url{http://cadcwww.hia.nrc.ca/cgps}.}.  The CGPS is supported by a grant from
the Natural Sciences and Engineering Research Council of Canada.

\clearpage

\begin{deluxetable}{rrrrrrrrrrrrrrrrl}
\rotate
\tabletypesize{\scriptsize} 
\setlength{\tabcolsep}{0.050in} 
\tablewidth{0pt}  
\tablecaption{Major CGPS HISA Features}
\tablehead{
 \colhead{Designation} & \multicolumn{3}{c}{$\dt$-Weighted Position\tablenotemark{a}} & \multicolumn{3}{c}{Coordinate Extent\tablenotemark{b}} & \colhead{Volume\tablenotemark{c}} & \colhead{Area\tablenotemark{d}} & \colhead{$\int{\!\dt}dV$\tablenotemark{e}} & \colhead{$\dtavg$\tablenotemark{f}} & \colhead{$\dt_{min}$\tablenotemark{g}} & \colhead{$\tuavg$\tablenotemark{h}} & \multicolumn{3}{c}{Diagnostics} & \colhead{Prior} \\
 \colhead{} & \colhead{$\glon$} & \colhead{$\glat$} & \colhead{$\vlsr$} & \colhead{$\Delta\glon$} & \colhead{$\Delta\glat$} & \colhead{$\Delta\vel$} & \colhead{} & \colhead{} & \colhead{} & \colhead{} & \colhead{} & \colhead{} & \colhead{$\tufrac$\tablenotemark{i}} & \colhead{$\fotfracvol$\tablenotemark{j}} & \colhead{$\fotfracint$\tablenotemark{k}} & \colhead{Detections\tablenotemark{l}} \\
 \colhead{} & \colhead{} & \colhead{} & \colhead{} & \colhead{} & \colhead{} & \colhead{} & \colhead{\ddeg$\,\cdot$~} & \colhead{} & \colhead{\kddeg$\,\cdot$~} & \colhead{} & \colhead{} & \colhead{} & \colhead{} & \colhead{} & \colhead{} & \colhead{} \\
 \colhead{} & \colhead{deg} & \colhead{deg} & \colhead{km/s} & \colhead{deg} & \colhead{deg} & \colhead{km/s} & \colhead{km/s} & \colhead{\ddeg} & \colhead{km/s} & \colhead{K} & \colhead{K} & \colhead{K} & \colhead{} & \colhead{} & \colhead{} & \colhead{Ref.\ ID}
}
\tablecolumns{17}
\startdata
GHISA 079.88$+$0.62$+$02& 79.882&$+0.620$&$+ 2.13$&11.075&7.710&26.38&54.407&16.671&$-764.83$&$-14.06$&$-68.66$&$105.03$&$0.98$&$0.90$&$0.94$&1,4,9,15,17,19,21,28\\
GHISA 083.87$+$1.74$-$74& 83.866&$+1.741$&$-74.41$& 0.480&0.370& 8.24& 0.160& 0.064&$  -1.31$&$ -8.20$&$-32.68$&$ 83.25$&$0.94$&$0.83$&$0.87$&\\                    
GHISA 085.25$+$0.54$-$45& 85.246&$+0.543$&$-45.14$& 0.670&0.535& 9.07& 0.412& 0.132&$  -5.86$&$-14.20$&$-37.96$&$ 85.77$&$0.91$&$0.90$&$0.92$&\\                    
GHISA 085.85$+$0.95$-$41& 85.852&$+0.949$&$-40.67$& 0.395&0.290& 6.60& 0.117& 0.051&$  -0.83$&$ -7.13$&$-27.59$&$ 85.69$&$1.00$&$0.84$&$0.88$&\\                    
GHISA 086.59$+$3.58$-$02& 86.589&$+3.576$&$- 1.89$& 0.690&0.630& 7.42& 0.368& 0.162&$  -3.69$&$-10.02$&$-38.12$&$101.37$&$1.00$&$0.80$&$0.84$&\\                    
GHISA 086.71$+$2.41$-$04& 86.710&$+2.412$&$- 4.25$& 0.325&0.205& 9.07& 0.111& 0.037&$  -1.24$&$-11.21$&$-36.35$&$102.18$&$1.00$&$0.85$&$0.91$&\\                    
GHISA 086.83$+$4.22$-$03& 86.833&$+4.216$&$- 2.97$& 0.545&0.575& 6.60& 0.222& 0.096&$  -2.05$&$ -9.26$&$-31.74$&$100.68$&$1.00$&$0.79$&$0.83$&\\                    
GHISA 088.09$+$4.72$-$02& 88.089&$+4.723$&$- 2.21$& 0.430&0.260& 8.24& 0.108& 0.054&$  -1.07$&$ -9.89$&$-27.80$&$ 90.86$&$1.00$&$0.80$&$0.82$&\\                    
GHISA 089.63$+$1.64$-$75& 89.632&$+1.639$&$-75.48$& 0.495&0.495& 9.89& 0.147& 0.058&$  -1.32$&$ -9.01$&$-29.69$&$ 79.61$&$0.52$&$0.78$&$0.82$&\\                    
GHISA 089.96$+$2.79$-$91& 89.963&$+2.791$&$-90.77$& 0.240&0.180& 7.42& 0.120& 0.031&$  -2.03$&$-16.87$&$-38.02$&$ 76.43$&$0.14$&$0.98$&$0.98$&\\                    
GHISA 090.01$+$2.10$-$86& 90.008&$+2.095$&$-86.27$& 0.555&0.640& 9.07& 0.468& 0.160&$  -7.83$&$-16.73$&$-40.83$&$ 73.14$&$0.01$&$0.95$&$0.96$&\\                    
GHISA 090.35$+$3.04$-$87& 90.352&$+3.044$&$-86.77$& 0.405&0.410& 9.07& 0.152& 0.066&$  -2.10$&$-13.82$&$-38.51$&$ 73.72$&$0.06$&$0.91$&$0.94$&\\                    
GHISA 090.66$+$2.38$-$82& 90.659&$+2.379$&$-82.21$& 0.480&0.650&10.72& 0.354& 0.120&$  -5.07$&$-14.33$&$-41.60$&$ 79.36$&$0.44$&$0.93$&$0.96$&\\                    
GHISA 091.39$+$1.91$-$82& 91.390&$+1.913$&$-81.77$& 0.330&0.430& 9.89& 0.197& 0.053&$  -4.43$&$-22.54$&$-52.52$&$ 76.54$&$0.17$&$0.96$&$0.98$&5\\                   
GHISA 091.63$+$2.00$-$72& 91.629&$+2.000$&$-71.87$& 0.365&0.315& 9.89& 0.139& 0.050&$  -1.38$&$ -9.87$&$-32.23$&$ 91.04$&$1.00$&$0.83$&$0.86$&\\                    
GHISA 091.90$+$3.27$-$03& 91.898&$+3.275$&$- 3.21$& 6.105&3.625&22.26&25.058& 7.543&$-601.78$&$-24.02$&$-80.72$&$ 88.20$&$0.74$&$0.95$&$0.98$&1,4,7,8,13,25,28\\    
GHISA 092.41$+$0.92$-$01& 92.411&$+0.923$&$- 1.20$& 0.400&0.465& 9.07& 0.162& 0.063&$  -1.73$&$-10.65$&$-37.26$&$ 99.06$&$1.00$&$0.83$&$0.87$&\\                    
GHISA 093.85$+$0.41$-$01& 93.853&$+0.412$&$- 0.55$& 0.990&1.435&11.54& 0.504& 0.247&$  -5.37$&$-10.65$&$-40.31$&$ 86.66$&$0.90$&$0.80$&$0.84$&\\                    
GHISA 094.32$-$0.10$-$43& 94.322&$-0.098$&$-43.27$& 3.205&2.260&18.96& 3.263& 1.274&$ -33.18$&$-10.17$&$-40.01$&$ 85.56$&$0.79$&$0.84$&$0.88$&5\\                   
GHISA 094.42$+$1.21$-$58& 94.422&$+1.214$&$-57.69$& 0.405&0.365& 9.89& 0.113& 0.039&$  -1.94$&$-17.16$&$-36.56$&$ 77.74$&$0.23$&$0.88$&$0.90$&\\                    
GHISA 095.39$+$1.06$-$64& 95.391&$+1.057$&$-63.72$& 0.360&0.240& 7.42& 0.132& 0.042&$  -1.38$&$-10.43$&$-34.06$&$ 84.43$&$1.00$&$0.92$&$0.94$&\\                    
GHISA 096.40$-$0.61$-$40& 96.401&$-0.606$&$-40.23$& 0.585&0.390& 9.07& 0.181& 0.069&$  -2.40$&$-13.23$&$-34.30$&$ 82.92$&$0.73$&$0.91$&$0.92$&\\                    
GHISA 097.16$+$1.76$-$65& 97.159&$+1.760$&$-65.17$& 0.800&0.655& 9.89& 0.306& 0.139&$  -2.77$&$ -9.06$&$-32.29$&$ 91.78$&$1.00$&$0.76$&$0.80$&\\                    
GHISA 098.69$+$0.70$-$13& 98.685&$+0.704$&$-12.94$& 1.335&1.385& 9.07& 1.209& 0.543&$ -14.59$&$-12.07$&$-37.90$&$ 77.74$&$0.22$&$0.86$&$0.88$&\\                    
GHISA 104.11$+$1.15$-$54&104.112&$+1.151$&$-54.11$& 0.395&0.375& 9.07& 0.188& 0.065&$  -2.18$&$-11.61$&$-33.54$&$ 90.75$&$0.99$&$0.87$&$0.90$&\\                    
GHISA 104.44$+$0.52$-$54&104.435&$+0.516$&$-53.86$& 1.885&1.005&18.14& 1.464& 0.538&$ -15.27$&$-10.44$&$-44.31$&$ 87.42$&$0.82$&$0.83$&$0.87$&\\                    
GHISA 106.53$+$0.93$-$59&106.533&$+0.934$&$-58.91$& 0.390&0.400&14.02& 0.241& 0.085&$  -2.28$&$ -9.45$&$-31.56$&$ 73.54$&$0.00$&$0.88$&$0.90$&\\                    
GHISA 107.45$+$0.21$-$51&107.449&$+0.213$&$-50.61$& 0.605&0.385& 7.42& 0.309& 0.105&$  -3.68$&$-11.94$&$-48.71$&$ 95.81$&$1.00$&$0.89$&$0.93$&5\\                   
GHISA 109.42$-$1.32$-$48&109.425&$-1.322$&$-47.82$& 0.625&0.955&10.72& 0.443& 0.194&$  -3.74$&$ -8.45$&$-27.50$&$ 78.32$&$0.15$&$0.84$&$0.88$&\\                    
GHISA 110.27$-$0.33$-$51&110.268&$-0.329$&$-51.39$& 0.560&0.525& 9.89& 0.287& 0.113&$  -3.48$&$-12.12$&$-41.32$&$ 76.02$&$0.18$&$0.85$&$0.89$&3\\                   
GHISA 113.13$-$0.98$-$37&113.128&$-0.985$&$-36.62$& 0.425&0.225& 5.77& 0.132& 0.063&$  -1.46$&$-11.00$&$-31.59$&$ 77.78$&$0.32$&$0.92$&$0.94$&\\                    
GHISA 113.34$+$1.01$-$11&113.336&$+1.013$&$-11.16$& 0.680&0.420& 9.07& 0.285& 0.111&$  -2.93$&$-10.28$&$-33.48$&$ 73.05$&$0.00$&$0.87$&$0.90$&\\                    
GHISA 114.48$+$2.18$-$13&114.481&$+2.180$&$-12.66$& 0.755&0.355& 4.12& 0.103& 0.068&$  -1.38$&$-13.40$&$-30.61$&$ 72.40$&$0.00$&$0.78$&$0.81$&\\                    
GHISA 115.39$-$1.67$-$43&115.394&$-1.674$&$-42.89$& 0.550&0.305& 7.42& 0.180& 0.075&$  -2.43$&$-13.53$&$-39.03$&$ 75.10$&$0.15$&$0.88$&$0.92$&\\                    
GHISA 116.36$-$2.02$-$45&116.364&$-2.017$&$-45.30$& 0.445&0.225& 6.60& 0.105& 0.040&$  -1.27$&$-12.07$&$-29.60$&$ 73.03$&$0.00$&$0.88$&$0.91$&\\                    
GHISA 118.69$+$2.60$-$64&118.691&$+2.598$&$-64.04$& 0.380&0.275& 7.42& 0.131& 0.052&$  -1.89$&$-14.40$&$-34.30$&$ 71.79$&$0.00$&$0.92$&$0.94$&\\                    
GHISA 119.34$-$1.56$-$35&119.344&$-1.561$&$-35.18$& 0.735&0.880& 8.24& 0.629& 0.261&$  -6.81$&$-10.83$&$-40.41$&$ 73.64$&$0.01$&$0.88$&$0.92$&5\\                   
GHISA 120.80$+$0.60$-$51&120.804&$+0.603$&$-51.23$& 0.895&1.100& 9.07& 0.925& 0.332&$ -11.78$&$-12.73$&$-47.33$&$ 82.30$&$0.71$&$0.88$&$0.93$&\\                    
GHISA 120.87$-$0.70$-$43&120.874&$-0.699$&$-42.98$& 0.660&0.505& 8.24& 0.316& 0.128&$  -3.85$&$-12.20$&$-40.56$&$ 75.12$&$0.05$&$0.91$&$0.95$&\\                    
GHISA 121.14$-$0.19$-$52&121.141&$-0.187$&$-51.96$& 0.410&0.250& 6.60& 0.114& 0.048&$  -1.30$&$-11.37$&$-30.79$&$ 72.55$&$0.00$&$0.86$&$0.90$&\\                    
GHISA 122.84$+$1.80$-$57&122.843&$+1.800$&$-56.59$& 1.025&0.390& 9.07& 0.248& 0.122&$  -2.72$&$-10.97$&$-36.35$&$ 82.03$&$0.73$&$0.80$&$0.84$&\\                    
GHISA 122.87$+$2.24$-$60&122.870&$+2.238$&$-59.57$& 0.840&0.410& 9.07& 0.358& 0.133&$  -4.51$&$-12.57$&$-38.91$&$ 89.87$&$1.00$&$0.90$&$0.92$&\\                    
GHISA 124.71$+$0.29$-$44&124.706&$+0.294$&$-43.93$& 0.420&0.595& 9.07& 0.244& 0.087&$  -2.71$&$-11.11$&$-37.41$&$ 80.11$&$0.49$&$0.87$&$0.91$&\\                    
GHISA 124.81$+$1.04$-$57&124.810&$+1.045$&$-56.94$& 0.510&0.385& 8.24& 0.186& 0.072&$  -1.71$&$ -9.23$&$-33.63$&$ 78.87$&$0.36$&$0.85$&$0.89$&\\                    
GHISA 126.68$+$0.76$-$55&126.682&$+0.759$&$-54.52$& 0.395&0.285& 6.60& 0.104& 0.050&$  -1.13$&$-10.84$&$-30.70$&$ 71.70$&$0.00$&$0.81$&$0.84$&\\                    
GHISA 126.89$+$1.61$-$58&126.887&$+1.613$&$-57.75$& 0.345&0.395& 9.07& 0.217& 0.081&$  -2.75$&$-12.66$&$-41.63$&$ 74.68$&$0.00$&$0.90$&$0.93$&\\                    
GHISA 127.72$+$0.94$-$55&127.716&$+0.941$&$-54.90$& 0.450&0.320& 7.42& 0.105& 0.049&$  -1.02$&$ -9.66$&$-33.23$&$ 92.38$&$1.00$&$0.75$&$0.78$&\\                    
GHISA 127.86$-$0.31$-$43&127.859&$-0.312$&$-43.28$& 0.355&0.440& 4.95& 0.112& 0.055&$  -0.95$&$ -8.46$&$-27.65$&$ 71.63$&$0.00$&$0.80$&$0.84$&\\                    
GHISA 128.15$+$0.21$-$44&128.152&$+0.207$&$-44.37$& 0.385&0.215&10.72& 0.108& 0.038&$  -1.49$&$-13.86$&$-38.18$&$ 78.79$&$0.35$&$0.87$&$0.89$&\\                    
GHISA 128.35$+$0.01$-$44&128.349&$+0.009$&$-44.28$& 0.385&0.180&13.19& 0.123& 0.036&$  -1.95$&$-15.94$&$-47.30$&$ 73.82$&$0.00$&$0.89$&$0.91$&\\                    
GHISA 129.50$+$0.97$-$49&129.499&$+0.973$&$-49.16$& 0.675&0.975& 7.42& 0.335& 0.147&$  -3.33$&$ -9.94$&$-43.52$&$ 89.73$&$0.99$&$0.78$&$0.82$&\\                    
GHISA 130.37$+$1.30$-$46&130.365&$+1.300$&$-46.15$& 0.720&0.425& 5.77& 0.125& 0.070&$  -1.02$&$ -8.10$&$-30.88$&$ 77.70$&$0.06$&$0.68$&$0.71$&\\                    
GHISA 133.71$-$0.03$-$49&133.714&$-0.032$&$-48.63$& 4.790&3.145&23.91&10.778& 3.505&$-166.85$&$-15.48$&$-57.40$&$ 89.89$&$0.87$&$0.90$&$0.93$&5,12,14,18\\          
GHISA 136.42$+$0.02$-$36&136.422&$+0.016$&$-36.50$& 0.505&0.345& 7.42& 0.152& 0.066&$  -1.53$&$-10.04$&$-33.66$&$ 83.43$&$0.89$&$0.82$&$0.85$&\\                    
GHISA 136.83$+$0.92$-$38&136.830&$+0.923$&$-37.73$& 0.990&0.920&11.54& 0.571& 0.251&$  -5.33$&$ -9.33$&$-34.00$&$ 74.51$&$0.01$&$0.82$&$0.85$&18\\                  
GHISA 137.88$-$0.70$-$40&137.880&$-0.703$&$-39.96$& 0.390&0.325& 8.24& 0.121& 0.044&$  -1.24$&$-10.25$&$-39.37$&$ 98.89$&$1.00$&$0.85$&$0.90$&\\                    
GHISA 137.95$-$0.27$-$37&137.954&$-0.265$&$-36.92$& 0.610&0.415& 6.60& 0.129& 0.070&$  -0.99$&$ -7.69$&$-34.67$&$100.50$&$1.00$&$0.65$&$0.69$&\\                    
GHISA 139.01$+$0.96$-$40&139.007&$+0.964$&$-40.32$& 2.245&1.685&12.37& 3.591& 1.292&$ -45.93$&$-12.79$&$-44.13$&$ 91.52$&$0.99$&$0.89$&$0.92$&5,18,22,23,24\\       
GHISA 139.21$-$0.60$-$40&139.207&$-0.597$&$-40.13$& 0.765&1.050& 9.07& 0.691& 0.278&$  -7.93$&$-11.49$&$-42.82$&$ 99.91$&$1.00$&$0.86$&$0.90$&24\\                  
GHISA 140.55$+$0.49$-$45&140.550&$+0.495$&$-44.79$& 0.965&0.930& 9.07& 0.396& 0.207&$  -4.47$&$-11.28$&$-33.17$&$ 94.07$&$1.00$&$0.83$&$0.85$&18,24\\               
GHISA 140.72$+$1.48$-$40&140.722&$+1.478$&$-40.30$& 0.655&0.755& 8.24& 0.275& 0.116&$  -2.92$&$-10.62$&$-38.27$&$ 92.51$&$1.00$&$0.80$&$0.83$&22\\                  
GHISA 141.10$-$0.60$-$41&141.100&$-0.604$&$-40.63$& 1.455&1.035&11.54& 1.317& 0.514&$ -15.89$&$-12.06$&$-47.18$&$103.22$&$1.00$&$0.87$&$0.91$&24\\                  
GHISA 141.37$-$1.70$-$41&141.366&$-1.698$&$-41.01$& 0.290&0.420& 5.77& 0.139& 0.062&$  -1.26$&$ -9.07$&$-29.24$&$ 87.02$&$0.99$&$0.80$&$0.83$&\\                    
GHISA 141.62$+$0.60$-$42&141.623&$+0.603$&$-41.82$& 0.595&0.320& 5.77& 0.107& 0.059&$  -0.91$&$ -8.55$&$-36.71$&$102.46$&$1.00$&$0.64$&$0.68$&24\\                  
GHISA 142.61$-$0.18$-$41&142.607&$-0.179$&$-41.41$& 0.750&1.070& 7.42& 0.452& 0.228&$  -4.00$&$ -8.86$&$-36.87$&$107.54$&$1.00$&$0.73$&$0.77$&\\                    
GHISA 144.07$+$0.72$-$33&144.066&$+0.717$&$-33.13$& 0.565&0.420& 9.89& 0.166& 0.077&$  -1.61$&$ -9.71$&$-30.79$&$ 83.99$&$0.89$&$0.83$&$0.86$&\\                    
GHISA 144.52$-$1.23$-$31&144.523&$-1.227$&$-30.73$& 0.645&0.755& 9.89& 0.507& 0.204&$  -8.72$&$-17.19$&$-51.57$&$ 92.53$&$1.00$&$0.87$&$0.91$&\\                    
GHISA 145.01$+$0.46$-$35&145.013&$+0.460$&$-34.68$& 0.910&1.610&14.84& 0.993& 0.353&$ -14.88$&$-14.98$&$-54.69$&$ 89.14$&$0.86$&$0.87$&$0.91$&\\                    
GHISA 145.81$+$0.96$-$36&145.813&$+0.958$&$-36.12$& 0.750&0.485& 9.07& 0.266& 0.106&$  -2.94$&$-11.05$&$-43.67$&$ 84.15$&$0.91$&$0.83$&$0.88$&\\                    
\enddata
\label{Tab:major_features}
\tablenotetext{a}{Average feature coordinates, weighted by $\dt$}
\tablenotetext{b}{Dimensions of the $(\glon,\glat,\vel)$ box containing the feature}
\tablenotetext{c}{Volume occupied by $|\dt| > 0$ voxels}
\tablenotetext{d}{Projected area of feature volume}
\tablenotetext{e}{3-D integrated absorption of feature}
\tablenotetext{f}{Mean absorption per voxel}
\tablenotetext{g}{Maximum absorption in any voxel ($\dt_{min} = -|\dt|_{max}$)}
\tablenotetext{h}{Mean $\tu$ per voxel}
\tablenotetext{i}{Fraction of detected volume for which $\tu \ge 80$~K and detections are most complete, based on $\fid$}
\tablenotetext{j}{Fraction of detected volume that is probably reliable, based on $\fot(\glon,\glat,\vel)$}
\tablenotetext{k}{Fraction of detected 3-D integrated absorption that is probably reliable, based on $\fot(\glon,\glat,\vel)$}
\tablenotetext{l}{Prior detections of this feature in the literature, if any; see Table~\ref{Tab:previous_detections} for reference notes}
\end{deluxetable} 

\clearpage

\begin{deluxetable}{rllllcrrr}
\tabletypesize{\scriptsize} 
\setlength{\tabcolsep}{0.04in} 
\tablewidth{0pt}  
\tablecaption{Prior HISA Studies in the CGPS Region}
\tablehead{
  \colhead{Ref.} & \colhead{Reference} & \colhead{Telescope} & \colhead{Beam\tablenotemark{a}} & \colhead{$\dv$\tablenotemark{b}} & \colhead{Map\tablenotemark{c}} & \multicolumn{3}{c}{Number of features} \\
  \colhead{ID} & \colhead{} & \colhead{} & \colhead{deg} & \colhead{km/s} & \colhead{} & \colhead{covered\tablenotemark{d}} & \colhead{found\tablenotemark{e}} & \colhead{major\tablenotemark{f}}
}
\tablecolumns{9}
\startdata
1 & \citet{d56} & Jodrell Bank 9m & 1.6 & 4.2 & n & 4 & 3 & 2 \\
2 & \citet{mg73} & Green Bank 91m\tablenotemark{g} & 0.22 & 1.0 & n & 1 & 1 & 0 \\
3 & \citet{hg73} & Green Bank 91m\tablenotemark{g} & 0.22 & 1.0 & Y & 1 & 1 & 1 \\
4 & \citet{k74} & Green Bank 43m & 0.35 & 0.08-0.34 & n & 4 & 3 & 2 \\
5 & \citet{ha74} & Green Bank 91m\tablenotemark{g} & 0.22 & 1.0 & n & 41 & 29 & 17 \\
6 & \citet{mg75} & Green Bank 91m\tablenotemark{g} & 0.22 & 1.0 & Y & 1 & 1 & 0 \\
7 & \citet{s76} & Dwingeloo 25m & 0.60 & 1.7 & Y & 1 & 1 & 1 \\
8 & \citet{r78} & Bonn 100m & 0.15 & 0.21 & n & 1 & 1 & 1 \\
9 & \citet{lrh80} & DRAO ST + 26m & 0.05 & 1.6 & Y & 1 & 1 & 1 \\
10 & \citet{r80} & Cambridge Half-Mile & 0.03 & 3.3 & Y & 4 & 2 & 0 \\
11 & \citet{s81} & Green Bank 43m & 0.35 & 0.69 & n & 1 & 0 & 0 \\
12 & \citet{r81} & Cambridge Half-Mile & 0.03 & 3.3 & Y & 4 & 4 & 4 \\
13 & \citet{sot81} & Hat Creek 26m\tablenotemark{h} & 0.59 & 1.1 & Y & 1 & 1 & 1 \\
14 & \citet{hsf83} & Green Bank 91m\tablenotemark{g} & 0.22 & 1.0 & Y & 6 & 6 & 6 \\
15 & \citet{wsd83} & Bonn 100m & 0.15 & 0.8 & Y & 1 & 1 & 1 \\
16 & \citet{m85} & Green Bank 91m\tablenotemark{g} & 0.22 & 1.0 & n & 3 & 1 & 0 \\
17 & \citet{stl88} & Green Bank 91m & 0.17 & 0.69 & n & 3 & 2 & 1 \\
18 & \citet{s90} & Green Bank 91m\tablenotemark{g} & 0.22 & 1.0 & Y & 13 & 10 & 8 \\
19 & \citet{f93} & DRAO ST + Bonn 100m & 0.03 & 1.3 & Y & 7 & 4 & 4 \\
20 & \citet{s95} & WSRT + Bonn 100m & 0.02 & 0.52  & Y & 1 & 1 & 0 \\
21 & \citet{ww96} & DRAO ST + Bonn 100m & 0.03 & 1.3 & n & 1 & 1 & 1 \\
22 & \citet{h99} & DRAO ST + 26m\tablenotemark{i} & 0.02 & 0.82 & Y & 3 & 3 & 3 \\
23 & \citet{g99} & DRAO ST + 26m\tablenotemark{i} & 0.02 & 0.82 & Y & 2 & 1 & 1 \\
24 & \citet{g00} & DRAO ST + 26m\tablenotemark{i} & 0.02 & 0.82 & Y & 3 & 3 & 2 \\
25 & \citet{ht00} & DRAO 26m & 0.62 & 0.82 & Y & 1 & 1 & 1 \\
26 & \citet{kb01} & DRAO ST + 26m\tablenotemark{i} & 0.02 & 0.82 & Y & 1 & 0 & 0 \\
27 & \citet{frk04} & DRAO ST + 26m\tablenotemark{i} & 0.02 & 0.82 & Y & 1 & 1 & 0 \\
28 & \citet{h05} & DRAO 26m & 0.62 & 0.82 & Y & 3 & 3 & 3 \\
\enddata
\label{Tab:previous_detections}
\tablenotetext{a}{Angular resolution (FWHM)}
\tablenotetext{b}{Velocity channel separation}
\tablenotetext{c}{Were maps made of HISA features?}
\tablenotetext{d}{Features within the CGPS Phase I coverage}
\tablenotetext{e}{Features detected wholly or partially in this HISA search (e.g., in Figure~\ref{Fig:hisa_overview_map})}
\tablenotetext{f}{Features overlapping with major features in this survey (see Table~\ref{Tab:major_features})}
\tablenotetext{g}{Maryland-Green Bank Galactic 21cm Line Survey data \citep{ww82}}
\tablenotetext{h}{Berkeley Low-Latitude Survey of Neutral Hydrogen data \citep{ww73}}
\tablenotetext{i}{Canadian Galactic Plane Survey data \citep{cgps}}
\end{deluxetable} 

\clearpage

\clearpage

\begin{figure}
\figurenum{1b}
\plotone{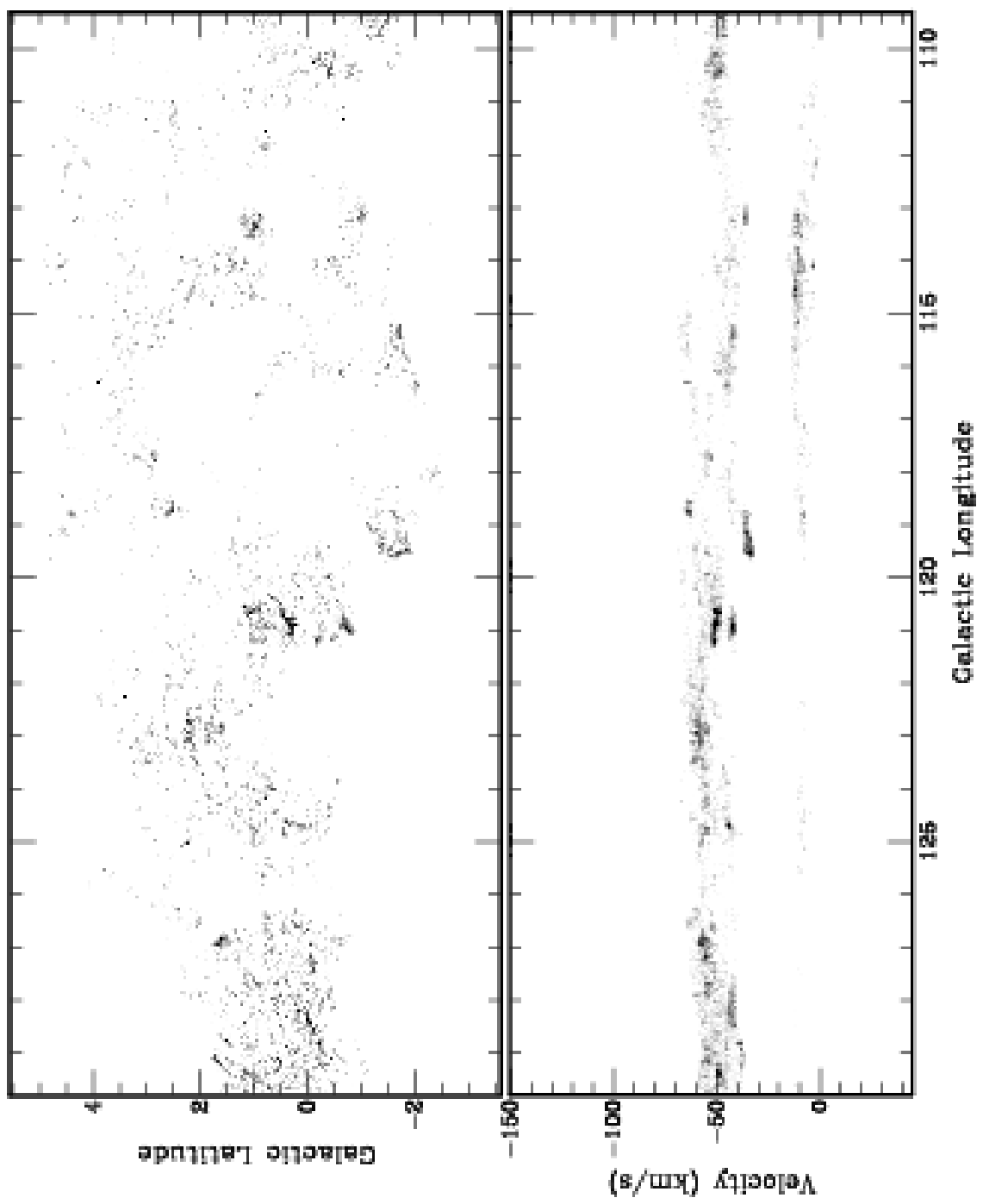}
\caption[]{
$(\glon,\glat)$ and $(\glon,\vel)$ maps of HISA $\dt$, continued.
}
\end{figure}

\clearpage

\begin{figure}
\figurenum{1c}
\plotone{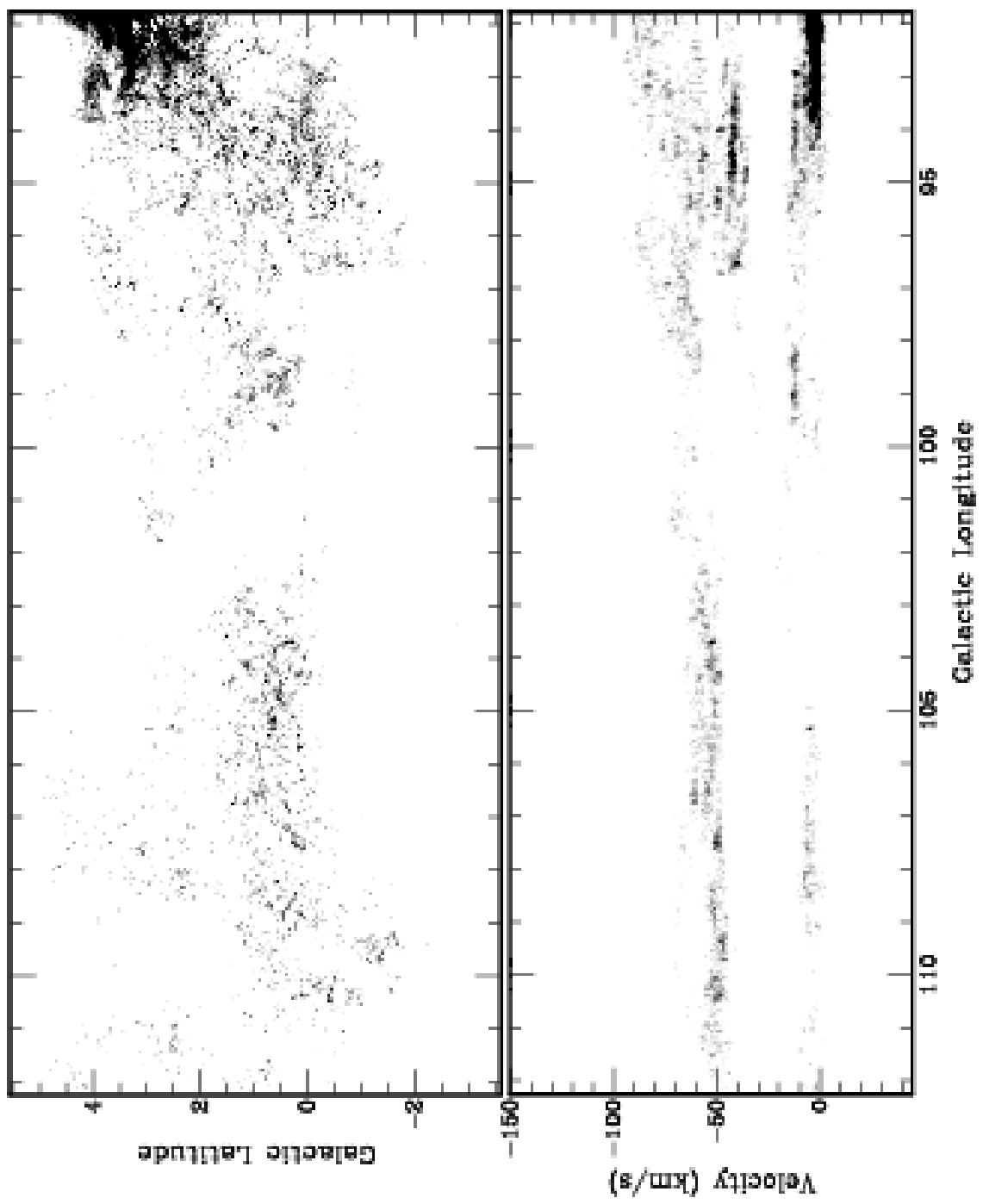}
\caption[]{
$(\glon,\glat)$ and $(\glon,\vel)$ maps of HISA $\dt$, continued.
}
\end{figure}

\clearpage

\begin{figure}
\figurenum{1d}
\plotone{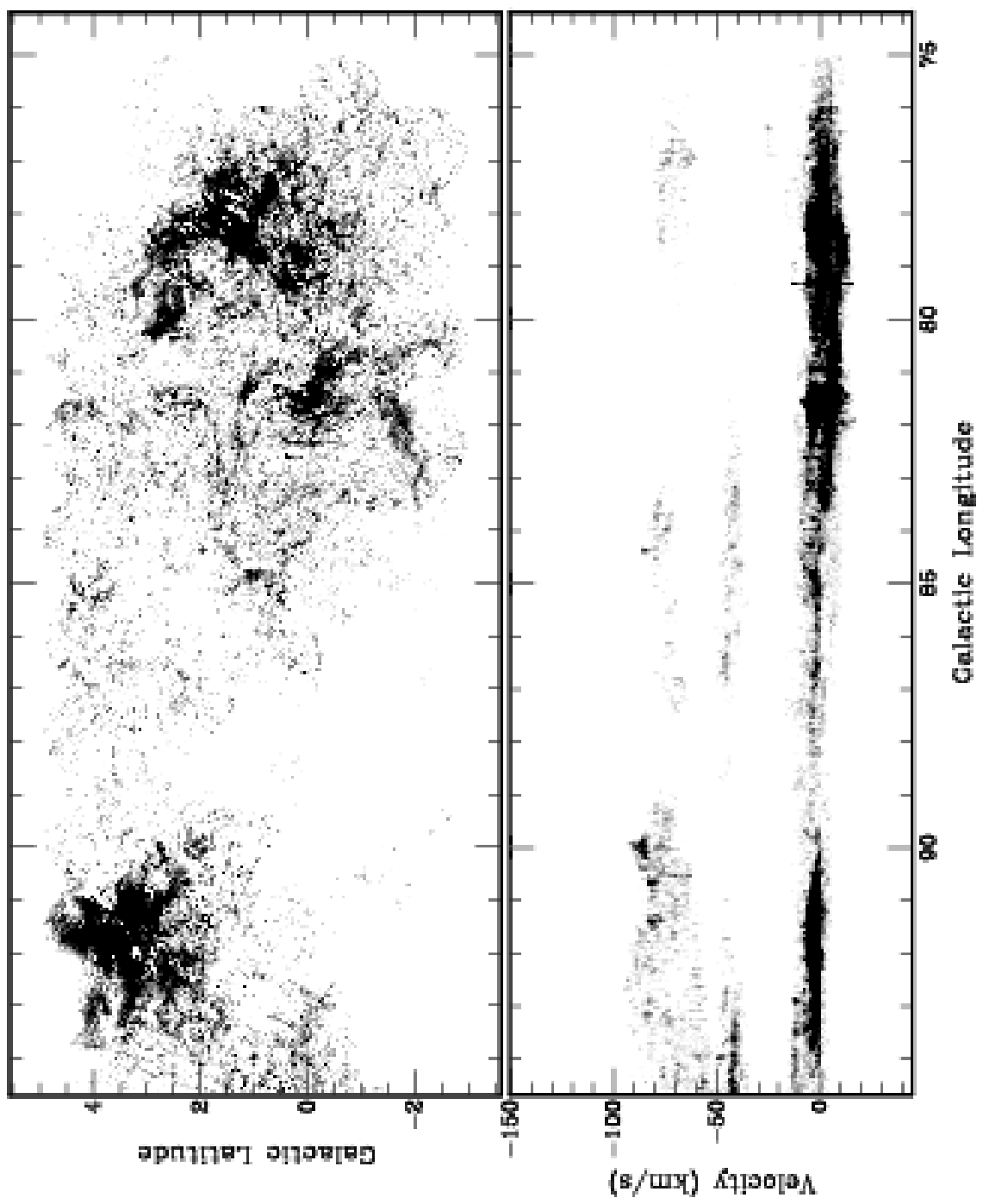}
\caption[]{
$(\glon,\glat)$ and $(\glon,\vel)$ maps of HISA $\dt$, continued.
}
\end{figure}

\clearpage

\begin{figure}
\plotone{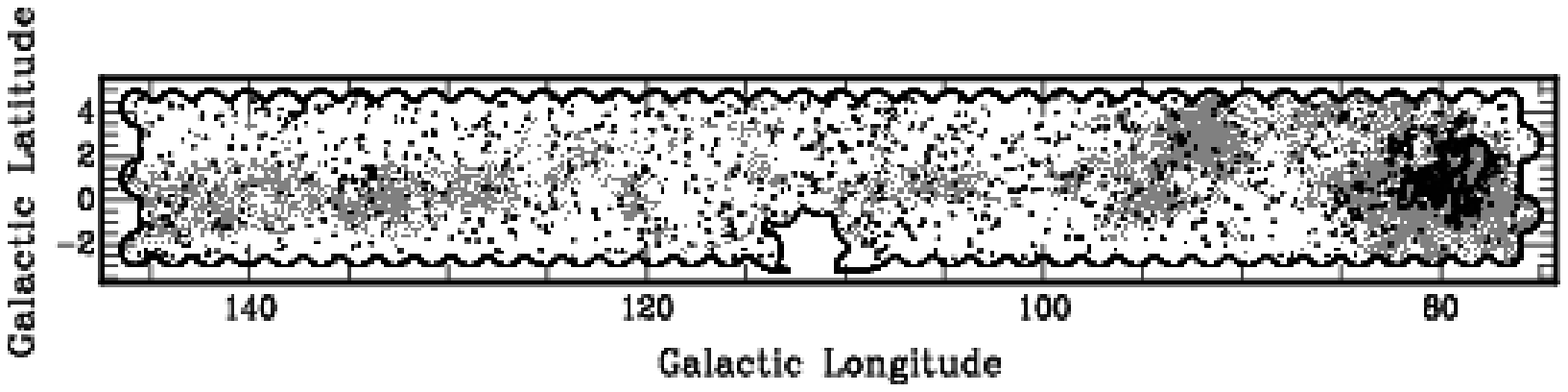}
\caption[]{
Limits on HISA detection from 20\arcmin\ mosaic edge clipping, synthesis field
weight clipping, and possible contamination from 21cm continuum emission.  The
background image shows all HISA detections over the $(\glon,\glat)$ area of the
survey (see Fig.~\ref{Fig:hisa_overview_map} for a more detailed view).  
Contours around the periphery indicate the boundaries set by the two clipping
requirements.  Additional contours scattered throughout the interior of the
survey outline regions where the continuum brightness $\tc$ is $\ge 20$~K, 
using matching-beam continuum maps provided by S. Strasser.  See text for 
discussion.
}
\label{Fig:survey_edges_and_c21}
\end{figure}

\begin{figure}
\plotone{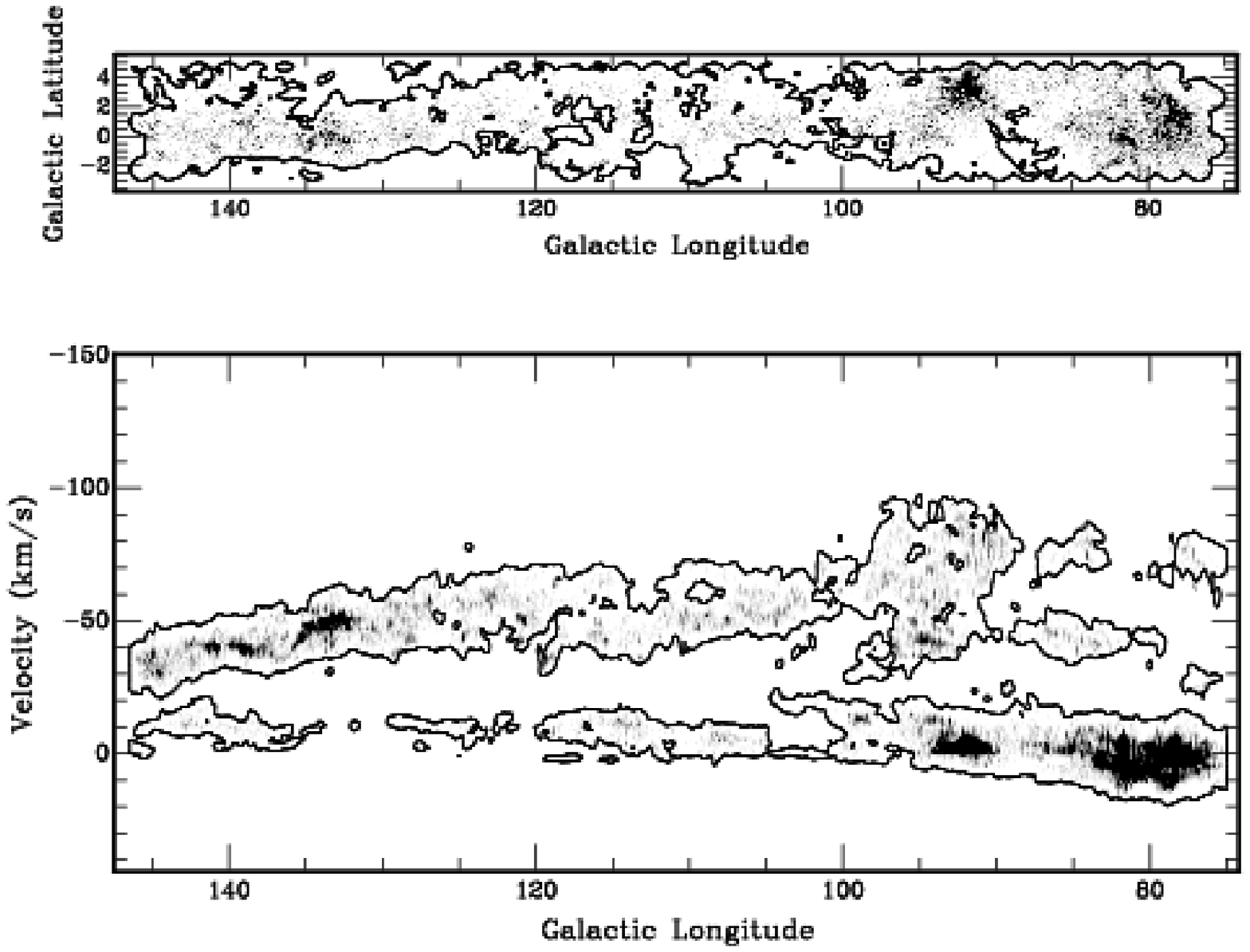}
\caption[]{
Detected HISA $\dt$ (background image) in $(\glon,\glat)$ and $(\glon,\vel)$, 
with the same intensity ranges used in Figure~\ref{Fig:hisa_overview_map}.
Contours show the approximate detection boundaries imposed by requiring 
$\tu \ge 70$~K in addition to the peripheral limits shown in 
Figure~\ref{Fig:survey_edges_and_c21}; these boundaries represent the 
``searchable area'' of the survey.  See text for further details.
}
\label{Fig:hisa_vs_hie}
\end{figure}

\begin{figure}
\plotone{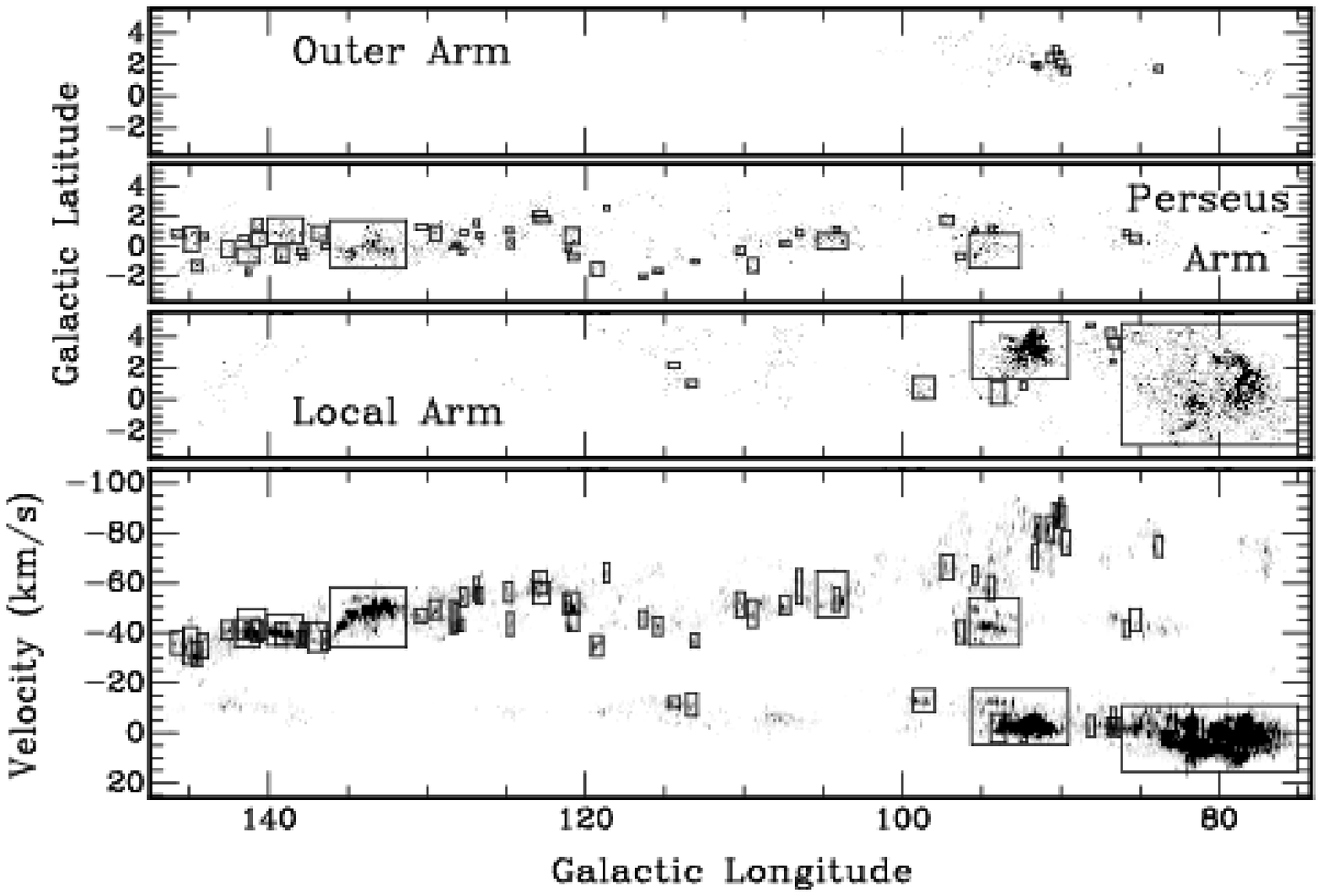}
\caption[]{
Detected HISA $\dt$ as shown in Figures~\ref{Fig:hisa_overview_map} \&
\ref{Fig:hisa_vs_hie}, but with the $(\glon,\glat)$ projections over the
velocity subranges $\vlsr < -70\kms$, $-20\kms < \vlsr -70\kms$, and $\vlsr >
-20\kms$, corresponding to the Outer, Perseus, and Local spiral arms,
respectively.  
Boxes mark the $(\glon,\glat,\vel)$ extents of features catalogued in 
Table~\ref{Tab:major_features}.
}
\label{Fig:major_features}
\end{figure}

\begin{figure}
\plotone{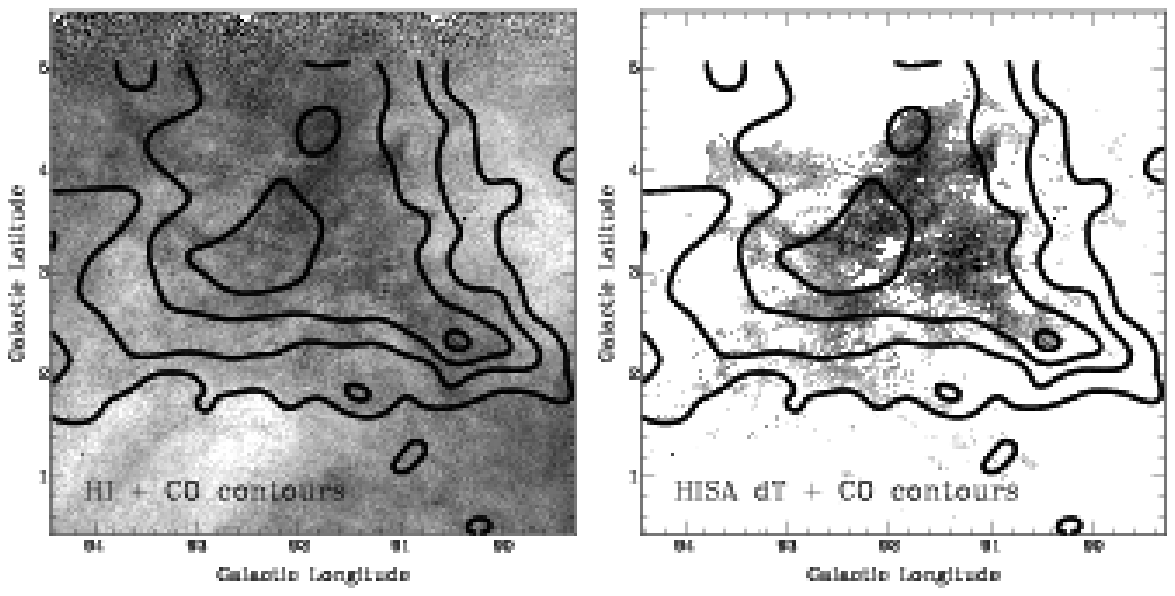}
\caption[]{
$(\glon,\glat)$ channel maps of the central concentration of feature
GHISA 091.90$+$3.27$-$03 at $-3$~\kms.  Shown are \hi\ 
brightness and HISA $\dt$, with \twco\ $J=1-0$ emission from \citet{d01}
at contour levels of 1, 2, 3, and 4~K.
Intensity ranges are 0 to $+130$~K for the \hi\ panel and $-65$ to 0~K 
for the $\dt$ panel.
The small white $6\arcmin \times 6\arcmin$ box 
$(\glon=91.15\arcdeg,\glat=+2.85\arcdeg)$
marks the area from which the 
spectra in Figure~\ref{Fig:real_hisa_id_stage_spectra} were taken.
The feature extraction is truncated for $\glat \ga +4.5$\arcdeg\ due to 
$\tu < 70$~K (see \S\ref{Sec:maps_overview} and Fig.~\ref{Fig:hisa_vs_hie}).
}
\label{Fig:real_hisa_id_stage_maps}
\end{figure}

\begin{figure}
\plotone{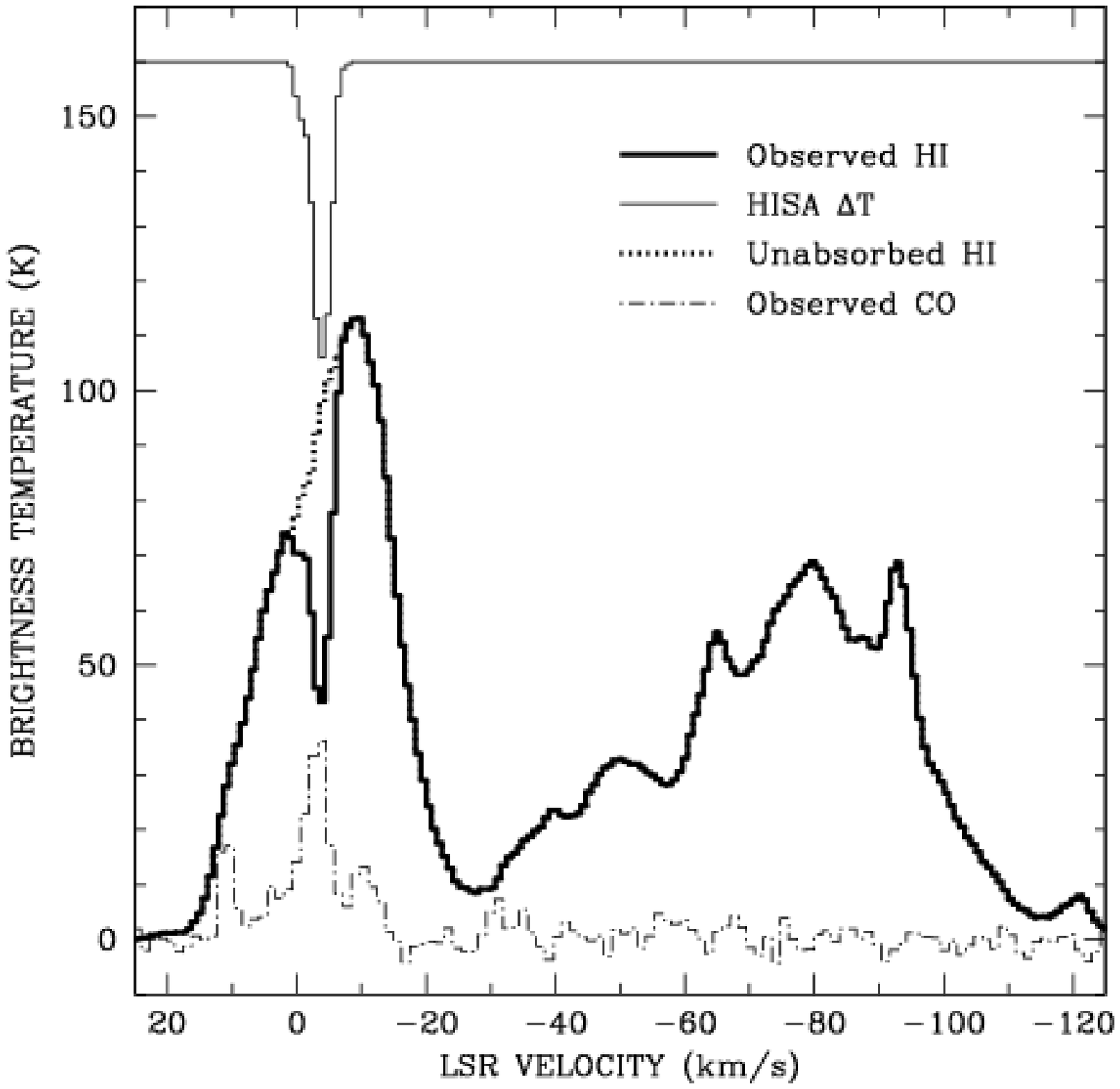}
\caption[]{
Spatially-averaged velocity spectra
extracted from the $6\arcmin \times 6\arcmin$ box marked in 
Figure~\ref{Fig:real_hisa_id_stage_maps} at 
$(\glon=91.15\arcdeg, \glat=+2.85\arcdeg)$.
Shown are \hi\ emission, HISA $\dt$, 
and \twco\ $J=1-0$ emission from \citet{d01}.
For clarity, the $\dt$ zero-point has been shifted to 160~K, and the CO 
brightness has been multiplied by 10.
}
\label{Fig:real_hisa_id_stage_spectra}
\end{figure}

\clearpage

\begin{figure}
\plotone{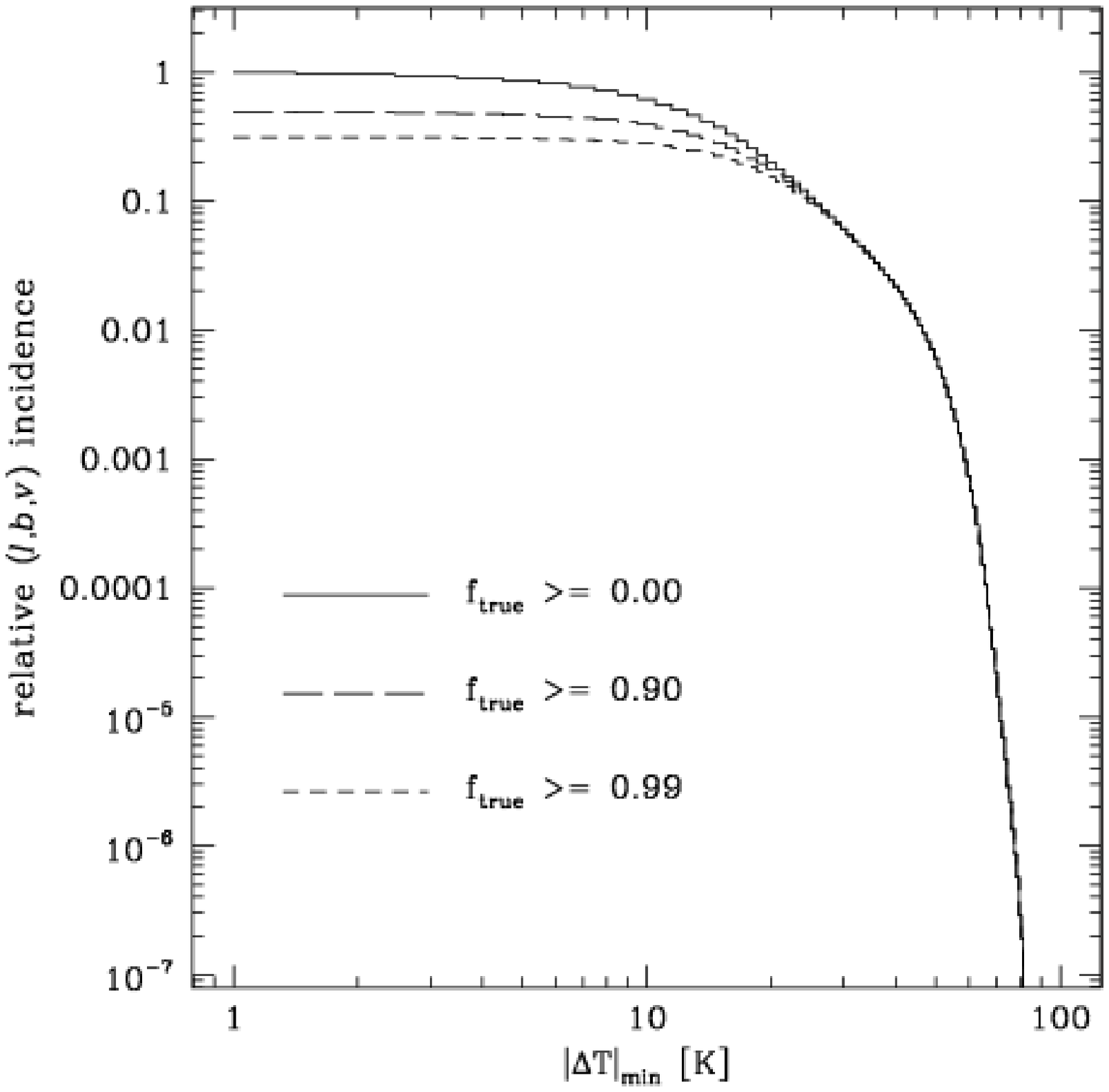}
\caption[]{
The relative incidence of HISA voxels in the CGPS for a minimum $|\dt|$ and a
minimum $\fot$.  49\% of all voxels have at least 90\% reliability, and 31\% 
have at least 99\% reliability.  This cumulative plot peaks at $|\dt|_{min} 
\rightarrow 0$, but few voxels are this faint; the $\dt$ distribution itself
peaks at $\sim -10$~K (see Paper~III, Fig.~11).
}
\label{Fig:hisa_coverage}
\end{figure}

\begin{figure}
\plotone{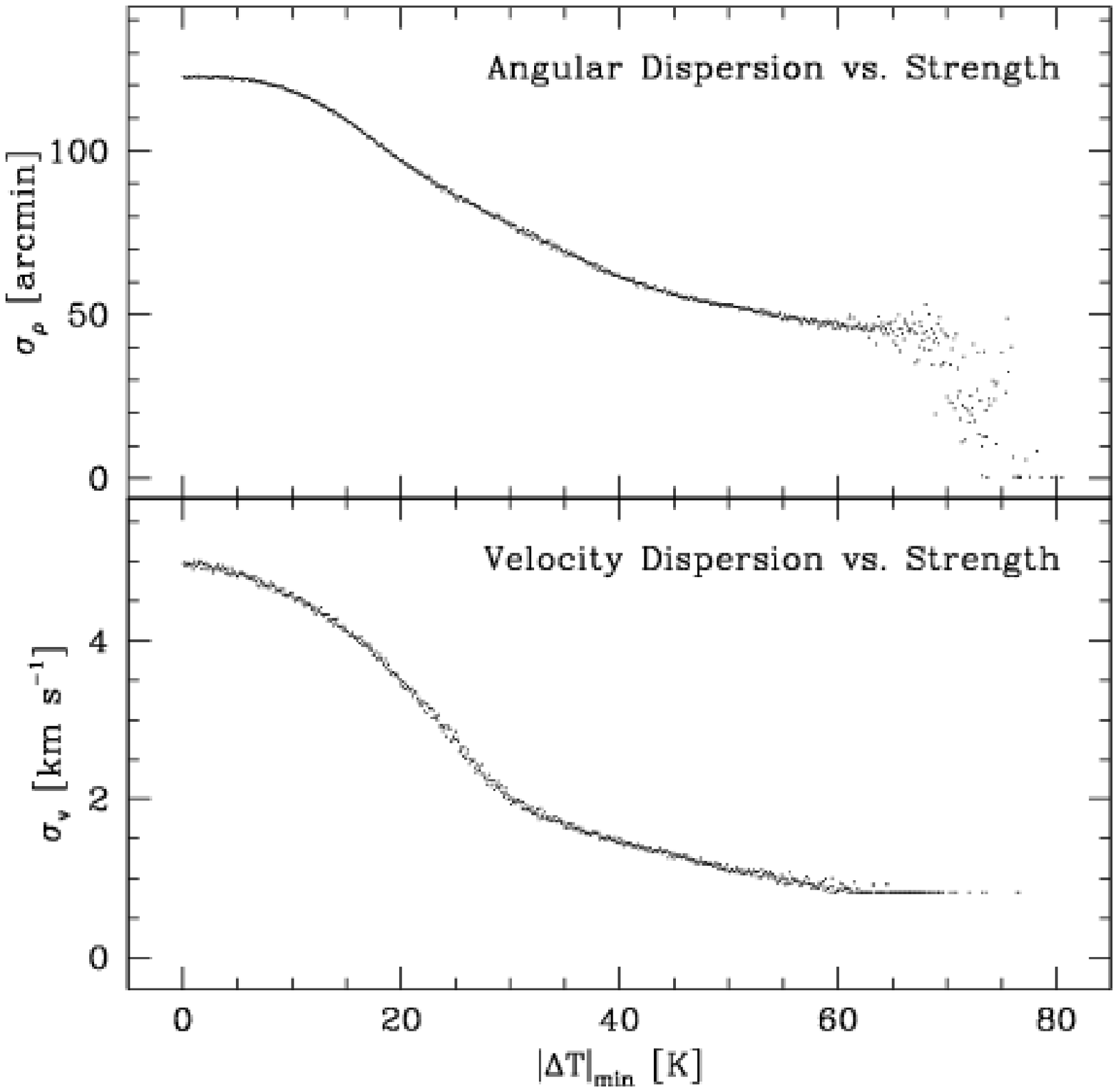}
\caption[]{
The dispersion of HISA voxel pairs in angle and velocity for HISA with
$|\dt| \ge |\dt|_{min}$ in both voxels.  Large-amplitude HISA is relatively 
more concentrated than small-amplitude HISA.
}
\label{Fig:strength_vs_dispersion}
\end{figure}

\begin{figure}
\plotone{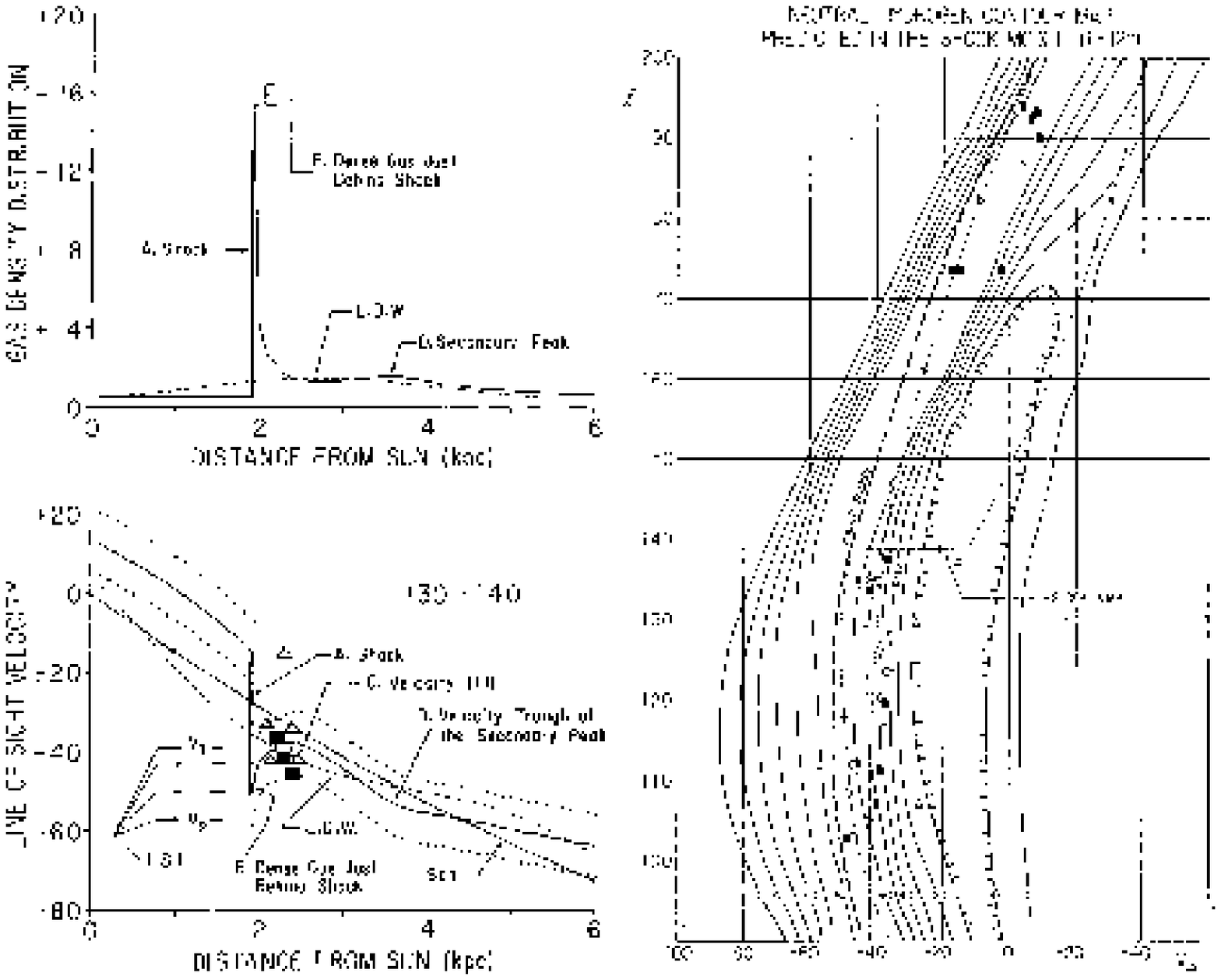}
\caption[]{
Plots from \citeauthor{r72b}'
(\citeyear{r72b}) Figures 3 and 6 that illustrate his two-arm spiral shock
(TASS) model, reproduced here with permission.
{\it Left:\/} Gas density and LSR velocity vs.\ distance for a 12\arcdeg\ 
pitch angle and $130\arcdeg \le \glon \le 140\arcdeg$.  
The observer is at left; gas flows from left to right, overtaking the Perseus
arm from behind.  The smooth solid line
is from an equilibrium \citet{s65} rotation model.  The dashed line shows a
linear density wave model, and the solid line with the discontinuity at $\sim
2$~kpc is the TASS model, with a $\pm 8$\kms\ dispersion band marked with
dotted lines.  The symbols mark various spiral arm tracers.
{\it Right:\/} 
$(\glon,\vel)$ plot of predicted gas emission brightness temperature for the 
same TASS model.  The dashed line marks the ``shock ridge'' (region~B in the 
lower-left panel), where dense gas occurs just behind the shock.
For full details, see \citet{r72b}.  
}
\label{Fig:roberts_shock}
\end{figure}

\begin{figure}
\plotone{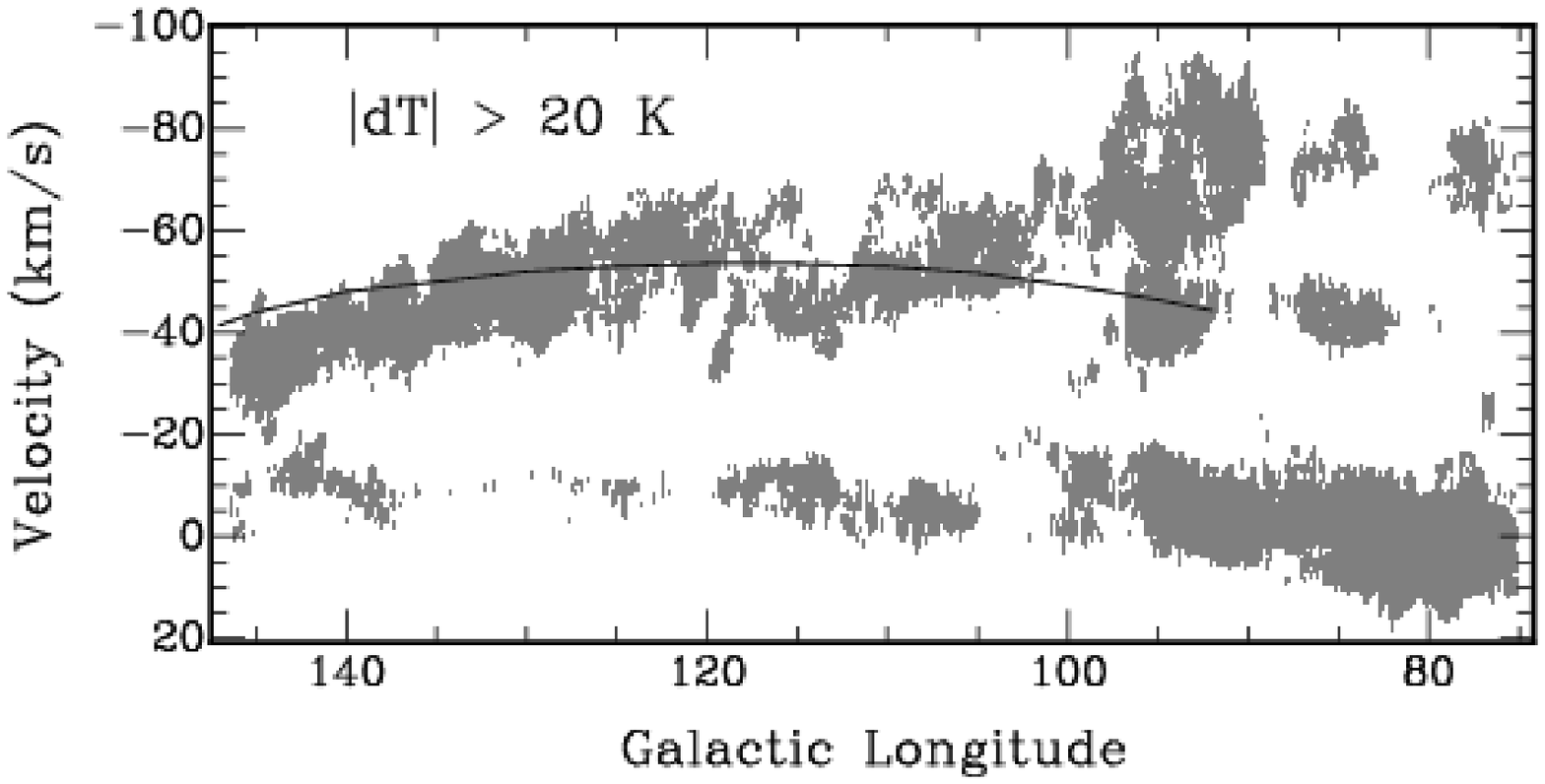}
\caption[]{
$(\glon,\vel)$ distributions of CGPS HISA projected over all $\glat$ for
a minimum voxel amplitude of $|\dt|_{min} = $ 20~K.  
The Perseus arm spiral shock ridge of \citet{r72b} is marked over all
longitudes that overlap with the CGPS (see Fig.~\ref{Fig:roberts_shock}).
}
\label{Fig:hisa_lv_shock}
\end{figure}

\end{document}